%% file: main.tex
\documentclass[sigconf,nonacm, pdfa]{acmart}

\usepackage[a-2b]{pdfx}
\usepackage{algorithm}
\usepackage{algorithmic}
\usepackage{multirow}
\newcommand{\up}[1]{{\uparrow \hspace{-3 pt} \ell_\textbf{#1}}}  
\newcommand{\down}[1]{{\downarrow \hspace{-3 pt} \ell_\textbf{#1}}}  
\newcommand{\upp}[1]{{\uparrow \hspace{-1 pt} \ell_\textbf{#1}}}  
\newcommand{\downn}[1]{{\downarrow \hspace{-1 pt} \ell_\textbf{#1}}}  
\newcommand{\downagg}{{\downarrow \hspace{-1 pt} \ell_\textrm{agg}}}
\newcommand{\upagg}{{\uparrow \hspace{-1 pt} \ell_\textrm{agg}}}
\newcommand{\EinSum}{{\texttt{EinSum}}}

\usepackage{listings}
\usepackage{amsmath}
\usepackage{enumitem}
\lstnewenvironment{SQL}
  {\lstset{
        aboveskip=5pt,
        belowskip=5pt,
        escapechar=!,
        mathescape=true,
        language=SQL,
        basicstyle=\linespread{0.94}\ttfamily\small,
        morekeywords={BULK, COLLECT, FOR, MATRIX, VECTOR, EACH, INTEGER, LONG, DOUBLE, WITH, PARTITION, TEST, WHETHER, PROBABILITY, VECTORIZE, ROWMATRIX, inner\_product, matrix\_vector\_multiply, MATRIX\_MULTIPLY, outer\_product, matrix\_inverse, trans\_matrix,
        label\_scalar, label\_vector, get\_scalar, diag, rowMins, rowIndexMax, map, transpose, multiply, asInstanceOf, DenseMatrix, reduce,
        toArray, zipped, toIndexedRowMatrix, if, Tuple2, override, def, new, gemm, build, filter, matmul, crossentropyderiv, reluderiv, t, relu, softmax, reducebyrow},
        deletekeywords={VALUE, PRIOR},
        showstringspaces=true}
        \vspace{0pt}%
        \noindent\minipage{0.47\textwidth}}
  {\endminipage\vspace{0pt}}

\newcommand\vldbdoi{10.14778/3797919.3797921}
\newcommand\vldbpages{1101 - 1114}
\newcommand\vldbvolume{19}
\newcommand\vldbissue{6}
\newcommand\vldbyear{2026}
\newcommand\vldbauthors{\authors}
\newcommand\vldbtitle{\shorttitle}
\newcommand\vldbavailabilityurl{https://github.com/yuxineverforever/upper-case-lower-case-einstein-notation}
\newcommand\vldbpagestyle{empty}

\begin{document}
\title{Automated Tensor-Relational Decomposition for Large-Scale Sparse Tensor Computation}

\author{Yuxin Tang, Zhiyuan Xin, Zhimin Ding, Xinyu Yao, Daniel Bourgeois, Tirthak Patel, Chris Jermaine}
\affiliation{%
  \institution{Rice University}
  \city{Houston}
  \state{Texas}
  \country{USA}
}
\email{ {yt33, zx58, zd21, xy38, dcb10, tp53, cmj4}@rice.edu}

\begin{abstract}
A \emph{tensor-relational} computation is a relational computation where individual tuples carry vectors, matrices, or higher-dimensional arrays.  An advantage of tensor-relational computation is that the overall computation can be executed on top of a relational system, inheriting the system's ability to automatically handle very large inputs with high levels of sparsity while high-performance kernels (such as optimized matrix-matrix multiplication codes) can be used to perform most of the underlying mathematical operations.  In this paper, we introduce upper-case-lower-case \texttt{EinSum}, which is a tensor-relational version of the classical Einstein Summation Notation. We study how to automatically rewrite a computation in Einstein Notation into upper-case-lower-case \texttt{EinSum} so that computationally intensive components are executed using efficient numerical kernels, while sparsity is managed relationally.
\end{abstract}

\maketitle

\pagestyle{\vldbpagestyle}
\begingroup\small\noindent\raggedright\textbf{PVLDB Reference Format:}\\
\vldbauthors. \vldbtitle. PVLDB, \vldbvolume(\vldbissue): \vldbpages, \vldbyear.\\
\href{https://doi.org/\vldbdoi}{doi:\vldbdoi}
\endgroup
\begingroup
\renewcommand\thefootnote{}\footnote{\noindent
This work is licensed under the Creative Commons BY-NC-ND 4.0 International License. Visit \url{https://creativecommons.org/licenses/by-nc-nd/4.0/} to view a copy of this license. For any use beyond those covered by this license, obtain permission by emailing \href{mailto:info@vldb.org}{info@vldb.org}. Copyright is held by the owner/author(s). Publication rights licensed to the VLDB Endowment. \\
\raggedright Proceedings of the VLDB Endowment, Vol. \vldbvolume, No. \vldbissue\ %
ISSN 2150-8097. \\
\href{https://doi.org/\vldbdoi}{doi:\vldbdoi} \\
}\addtocounter{footnote}{-1}\endgroup

\ifdefempty{\vldbavailabilityurl}{}{
\vspace{.3cm}
\begingroup\small\noindent\raggedright\textbf{PVLDB Artifact Availability:}\\
The source code, data, and/or other artifacts have been made available at \url{\vldbavailabilityurl}.
\endgroup
}

\section{Introduction}

In machine learning, a tensor is a specialized relation, mapping a key specifying a position in a multidimensional array to a scalar.  Given the close relationship between tensors and relations, almost all computations over tensors can be implemented on top of a relational database system.  

For example, consider the matrix multiplication chain $(\textbf{X} \times \textbf{Y})^T \times \textbf{Z}$.  This multiplication is used to implement a message passing scheme in a graph neural network \cite{wu2020comprehensive,zhou2020graph}.  That is, each column in \textbf{X} stores an embedding for a vertex, \textbf{Y} encodes the (directional) edges between vertices, and the goal is to sum all embeddings being passed over edges to a given vertex, and then to compute a linear transformation (represented by \textbf{Z}) over the sums at each vertex.  
If all input matrices are stored as \texttt{(row, col, value)} triples, the SQL corresponding to this computation is as follows:

\begin{SQL}
SELECT XtimesY.row, Z.col, 
       SUM (XtimesY.value * Z.value) AS value 
FROM (SELECT X.row AS col, Y.col AS row, 
             SUM (X.value * Y.value) AS value 
      FROM X, Y WHERE X.col = Y.row
      GROUP BY X.row, Y.col) AS XtimesY, Z
WHERE XtimesY.col = Z.row
GROUP BY XtimesY.row, Z.col
\end{SQL}

\noindent
\textbf{Is the relational implementation effective?} Unfortunately, this computation is unlikely to run well on a relational system.  Imagine that a large, sparse graph represented by \textbf{Y} has 100 million vertices, each with 1000 neighbors, on average.  The embeddings of each vertex (stored in \textbf{X}) are of size 8192, and the transformation \textbf{Z} results in a size 8192 vector.  In this case, the first join of  \textbf{X} and  \textbf{Y} will produce approximately 820 trillion intermediate tuples, which are then aggregated down to 820 billion tuples.  This result is then joined with \textbf{Z}, resulting in \emph{6,000 trillion tuples}, a debilitating number.

One could instead run this computation directly using tensors on top of a conventional deep learning system (for example, PyTorch \cite{pytorch}).  This would facilitate use of accelerators such as GPU, but the results are likely to be unsatisfactory.  There is the problem that storing $\textbf{X}$ requires 3.2TB of GPU RAM (at half precision). This requires 40 GPUs (assuming 80GB of storage each). Then there is the problem of computation. GPUs are not meant for sparse computation, and for very sparse matrices, even state-of-the-art, hand-tuned sparse-dense GPU kernels operate at around 0.1\% compute utilization on an A100 GPU \cite{xiang2025cutespmm}.  At 0.1\% compute utilization, the $1.6 \times 10^{15}$ floating point operations required for the first matrix multiply would require more than an hour of A100 GPU time. 

\vspace{5 pt}
\noindent
\textbf{A sparse tensor-relational implementation.}
Instead, one could run this computation \emph{tensor-relationally} \cite{Tensorrelationalalgebra}, decomposing the problem so that a relational system can take advantage of the sparsity inherent in the computation, but efficient dense CPU (or GPU) kernels can be used where appropriate:

\begin{SQL}
X (col INT, value VECTOR[8192])
Y (row INT, col INT, value DOUBLE)
Z (value MATRIX[8192, 8192])
\end{SQL}

\begin{SQL}
SELECT XtimesY.row, 
    vec_mat_mult (XtimesY.value, Z.value) AS value
FROM (SELECT Y.col AS row, 
      SUM (vec_sca_mult (X.value, Y.value)) AS value 
      FROM X, Y WHERE X.col = Y.row
      GROUP BY Y.col) AS XtimesY, Z
\end{SQL}

Kernels (\texttt{vec\_mat\_mult()} and \texttt{vec\_sca\_mult()}) will be very efficient, and because tuples are grouped into \texttt{VECTOR} and \texttt{MATRIX} structures, the number of intermediate tuples is radically reduced.  The join of \textbf{X} and \textbf{Y} will now produce only 100 billion tuples, and the second join will produce only 100 billion tuples. 

Further, by running this computation on top of a relational system, memory and parallelization/distribution will be automatically managed.  A distributed database system can automatically parallelize the computation across multiple machines, managing memory so that the ``Out-Of-Memory'' errors ubiquitous in modern machine learning programming are avoided.

\vspace{5 pt} \noindent \textbf{Sparse tensor-relational decompositions via upper-case-low\-er-case \texttt{EinSum}.}  In the above example, the choice was made to decompose \textbf{X} into row vectors, \textbf{Y} into scalars, and to leave \textbf{Z} un-decomposed.  The question we address in this paper is: how can we take an arbitrary tensor computation and decompose it into a tensor-relational computation where computationally intensive portions of the computation are handled using efficient kernels (such as \texttt{vec\_mat\_mult()}) but where sparsity is handled relationally?  

To address this challenge, we propose a new tensor-relational variant of Einstein summation notation \cite{barr1991einstein}, referred to as upper-case–lower-case \texttt{EinSum}. Classic \texttt{EinSum} is a standard tensor calculus used to express machine learning and numerical computations. In upper-case–lower-case \texttt{EinSum}, expressions explicitly specify which parts of the computation are handled relationally and which parts are executed using efficient numerical kernels.

Given a directed acyclic graph of \texttt{EinSum} computations, the task of producing an equivalent, optimized tensor-relational computation can be reduced to automatically rewriting expressions into equivalent upper-case–lower-case \texttt{EinSum} expressions that maximize performance. In this paper, we propose an algorithm to perform this rewrite, called \textsc{SparseEinSum}. The \textsc{SparseEinSum} algorithm combines a simple cost model that evaluates the quality of an upper-case–lower-case \texttt{EinSum} expression under sparsity with a dynamic programming approach to optimize the rewrite.

We demonstrate experimentally that \textsc{SparseEinSum} produces high-performing tensor-relational computations. Our experimental testbed includes a variety of sparse tensor workloads, such as large-scale graph neural networks, graph-based attention computations, and quantum circuit simulation.

\section{The upper-case-lower-case \EinSum{} Approach}

Assume we are given a (potentially) complex numerical computation specified as a directed acyclic graph (DAG) of \texttt{EinSum} expressions, and our goal is to produce an optimized computation that can be executed on top of a relational system.

\texttt{EinSum} is described in detail in Section 3. At a high level, it asks the programmer to specify a simple declarative expression describing how the entries of an output tensor are computed. For example, consider the matrix multiplication $\mathbf{W} \leftarrow \mathbf{X} \times \mathbf{Y}$. In \texttt{EinSum}, this computation can be expressed as:
\begin{align}
\forall \hspace{2 pt} i,k \hspace{4 pt} \textbf{W}_{i,k} \leftarrow
\sum_{j}  
  \textbf{U}_{i,j} \times \textbf{V}_{j,k} \noindent
  \label{eqn-Einsum}
\end{align}
By allowing arbitrary aggregation and scalar functions, as well as arbitrary indexing into the input tensors, the output tensor, and the aggregation domain, \texttt{EinSum} can express a very broad class of tensor computations. As a result, \texttt{EinSum} is now widely used in machine learning. Popular frameworks such as NumPy \cite{numpy}, JAX \cite{jax}, PyTorch \cite{pytorch}, and Dask \cite{dask} all provide implementations of \texttt{EinSum}. If it is unreasonable to require programmers to write \texttt{EinSum} expressions directly, they may instead specify computations using a PyTorch-like syntax, in which high-level operators (e.g., \texttt{matmul}) act as thin wrappers around the underlying \texttt{EinSum} representation.

There is a close relationship between \texttt{EinSum} and SQL/relational databases, in that every \texttt{EinSum} expression can be translated into SQL and implemented as a sequence of joins followed by an aggregation \cite{EfficientandPortableSQL}. However, the standard \texttt{EinSum}-to-SQL translation implicitly assumes that all data are stored in fully relational form, with individual scalar values stored in tuples. If not handled carefully, such a decomposition can result in poor performance.

To this end, we introduce the concept of upper-case–lower-case \texttt{EinSum} notation, which serves as the target representation for our \textsc{SparseEinSum} rewrite. In this notation, indices written in upper case are handled relationally (i.e., they are “promoted”), whereas indices written in lower case are handled by tensor indexing (i.e., they are “demoted”). For example, consider the task of multiplying two $8192 \times 8192$ matrices. If we write Equation~\ref{eqn-Einsum} as:
\begin{align*}
\forall \hspace{2 pt} i,k \hspace{4 pt} \textbf{W}_{i,k} \leftarrow
\sum_{J}  
  \textbf{U}_{i,J} \times \textbf{V}_{J,k} \noindent
\end{align*}
\noindent it corresponds to the following, tensor-relational implementation:

\begin{SQL}
U (J INT, valU VECTOR[8192])
V (J INT, valV VECTOR[8192])

SELECT SUM (outer_prod (U.valU, V.valV)) AS valW
FROM U, V
WHERE U.J = V.J
\end{SQL}

\noindent Note that since $J$ is the only index in upper-case notation, it is the only one to appear in the SQL code. The indices $i$ and $k$ are handled implicitly by the \texttt{outer\_prod()} kernel function. The output tensor \textbf{W} is then stored as a single tuple storing the complete output 8192 $\times$ 8192 matrix. As an alternative, we can write:
\begin{align*}
\forall \hspace{2 pt} I,K \hspace{4 pt} \textbf{W}_{I,K} \leftarrow
\sum_{j}  
  \textbf{U}_{I,j} \times \textbf{V}_{j,K}
\end{align*}
This corresponds to the following:

\begin{SQL}
U (I INT, valU VECTOR[8192])
V (K INT, valV VECTOR[8192])

SELECT U.I, V.K, inner_prod (U.valU, V.valV) AS valW
FROM U, V
\end{SQL}

Again, since $I$ and $K$ are handled relationally and $j$ is not, $j$ does not appear anywhere in the corresponding SQL; it is handled by the kernel function \texttt{inner\_prod()} which loops over the two vectors \texttt{valU} and \texttt{valV} along the $j$ dimension. The output matrix \textbf{W} is stored as a set of scalars indexed by $I$ and $K$.

Given the upper-case-lower-case notation, the core problem then becomes: how to re-write a DAG of \texttt{EinSum} expressions into a DAG of upper-case-lower-case \texttt{EinSum} expressions that leverages a relational engine as appropriate to handle sparsity, and leverages high-performance kernels as appropriate to leverage density?

To this end, the remainder of the paper is organized as follows. The next section of the paper describes \texttt{EinSum} more formally.  Then Section \ref{sec-sparse} describes the upper-case-lower-case notation and how it can take advantage of sparsity. Section \ref{sec-compiling} describes how upper-case-lower-case \texttt{EinSum} expressions can be translated into SQL and thus executed on existing relational systems. Sections \ref{sec-costmodel} and \ref{sec-optimizing} describe the two major components of the \textsc{SparseEinSum} re-write algorithm: a cost model for predicting the utility of a re-write into upper-case-lower-case \texttt{EinSum}, and then a simple dynamic programming procedure for optimizing the cost of the re-write. Experiments in Section \ref{sec-experimeents} evaluate the utility of the approach.

\section{EinSum Background}
\input{ChrisText}

\begin{table}[t]
    \centering
    \begin{tabular}{|l|c|c|c|c|}
    \hline
    \textbf{Dataset} & \textbf{Classes} & \textbf{Features} & \textbf{Nodes} & \textbf{Edges} \\ \hline
    Cora & 7 & 1,433 & 2,485 & 5,069 \\ \hline
    Planetoid.Cora \cite{yang2016revisiting} & 7 & 1,433 & 2,708 & 10,556 \\ \hline
    Planetoid &  &  &  & \\
    .CiteSeer \cite{yang2016revisiting} & 6 & 3,703 & 3,327 & 9,104 \\ \hline
    CitationFull &  &  &  & \\
    .CiteSeer \cite{bojchevski2017deep} & 6 & 602 & 4,230 & 10,674 \\ \hline
    Amazon.Photo \cite{shchur2018pitfalls} & 8 & 745 & 7,650 & 0.2M \\ \hline
    Amazon &  &  &  & \\
    .Computers \cite{shchur2018pitfalls} & 10 & 767 & 13,752 & 0.4M \\ \hline
    ogbn-arxiv* \cite{hu2020open} & 40 & 128 & 169,343 & 1.1M \\ \hline
    ogbn-products* \cite{hu2020open} & 47 & 100 & 2.4M & 61.8M \\ \hline
    ogbn- &  &  &  & \\
    papers100M* \cite{hu2020open} & 172 & 128 & 111M & 1.6B \\ \hline
    friendster* \cite{boyd2004friendster} & 100 & 128 & 65M & 1.8B \\ \hline
    \end{tabular}
    \caption{Statistics of the datasets after graph standardization. Datasets marked with * are used for distributed execution.}
    \vspace{-15 pt}
    \label{table:graph-dataset-statistics}
\end{table}

\begin{table}[t!]
    \centering
    \setlength{\tabcolsep}{5.5pt} 
    \begin{tabular}{|c c|c|c|c|c|c|}
        \hline
        \multicolumn{2}{|c|}{\textbf{Dataset and}} &  \multicolumn{2}{c|}{\textbf{Tensor-based}} & \multicolumn{3}{c|}{ \textsc{SparseEinSum}} \\
        \cline{3-7}
        \multicolumn{2}{|c|}{\textbf{num machines}}  & \textsc{DGL} & \textsc{Ali} & \textbf{Opt} & \textbf{2nd} & \textbf{3rd} \\
        \hline
        \multirow{4}{*}{ogbn-arxiv} 
            & 1 & \textbf{4.3} & 12.9 & 8.3 & 13.4 & 21.8 \\
            & 2 & \textbf{1.6} & 6.9 & 5.2 & 8.8 & 12.6 \\
            & 4 & \textbf{0.9} & 3.4 & 2.0 & 4.1 & 6.3 \\
& 8 & \textbf{0.6} & 2.0 & 1.2 & 3.6 & 5.7 \\
        \hline
        \multirow{4}{*}{ogbn-products} 
            & 1 & \textbf{20.3} & 96.7 & 31.4 & 35.7 & 37.2 \\
            & 2 & 17.9 & 51.1 & \textbf{16.5} & 22.6 & 28.1 \\
            & 4 & 9.6 & 31.7 & \textbf{8.4} & 11.8 & 15.9 \\
            & 8 & 6.3 & 18.8 & \textbf{5.9} & 8.6 & 11.2 \\
        \hline
            & 1 & OOM & OOM & \textbf{289.5} & 306.1 & 351.9 \\
            ogbn- & 2 & OOM & OOM & 171.4 & \textbf{168.4}  & 207.4 \\
            papers100M & 4 & 92.5 & OOM & \textbf{82.1} & 93.5 & 95.1 \\
            & 8 & 59.3 & OOM & \textbf{42.4} & 52.0 & 61.1 \\
        \hline
        \multirow{4}{*}{friendster} 
            & 1 & OOM & OOM & \textbf{473.9} & 686.0 & 721.1 \\
            & 2 & OOM & OOM & \textbf{324.0} & 413.1 & 556.1 \\
            & 4 & OOM & OOM & \textbf{156.9} & 277.6 & 381.3 \\
            & 8 & 103.1 & OOM & \textbf{94.5} & 186.4 & 206.1 \\
        \hline
    \end{tabular}
    \caption{Running time comparison (in seconds) for one iteration of GCN training across different cluster sizes. DGL (PyTorch) and AliGraph are compared with the top three \textsc{SparseEinSum} decompositions.}
    \label{table:gcn_distributed_running_times}
    \vspace{-15pt}
\end{table}

\section{Experiments}
\label{sec-experimeents}

\subsection{Experimental Overview}

Experimentally, we wish to answer the question: \emph{Can upper-case-lower-case Einstein notation generated by \textsc{SparseEinSum} show better scalability over traditional tensor or purely relational approaches?}

\subsection{Evaluating \textsc{SparseEinSum} Decompositions}

\noindent
\textbf{Distributed graph convolution neural networks (GCNs) \cite{kipf2016semi}.} We compare \textsc{SparseEinSum}-generated tensor-relational computations with DGL v2.4.0 \footnote{2.4.x branch until commit c6c874b}\cite{wang2020deepgraphlibrarygraphcentric} running on a PyTorch backend \cite{pytorch}---this is a fully tensor-based system---as well as AliGraph \cite{zhu2019aligraphcomprehensivegraphneural}. These systems are chosen as they represent the most well-known systems for graph neural network computation. The distributed engine used to run the \textsc{SparseEinSum}-generated tensor-relational computation is PlinyCompute \footnote{\url{https://github.com/riceplinygroup/plinycompute}} \cite{zou2018plinycompute}, a distributed relational computation engine. We use TACO \footnote{\url{https://github.com/tensor-compiler/taco}} compiler-generated kernels \cite{taco,10.1145/3276493,kjolstad2019tensor} to efficiently support tensor operations of arbitrary shape within the tensor-relational computations generated by \textsc{SparseEinSum}.

\vspace{5 pt} 
\noindent
\textit{Computational problem.} The core computation in a graph convolutional neural network is given as the following \EinSum{}:
\begin{align}
    &\textbf{T}_{i,k}^{0} \leftarrow \sum_{j} \hat{\textbf{D}}_{i,j}\times \hat{\textbf{A}}_{j,k}; \hspace{20 pt}
\textbf{T}_{i,l}^{1} \leftarrow \sum_{k} \textbf{T}_{i,k}^{0}\times \hat{\textbf{D}}_{k,l} \nonumber \\
&\textbf{T}_{i,m}^{2} \leftarrow \sum_{l} \textbf{T}_{i,l}^{1} \times \textbf{H}_{l,m}^{(l)} ; \hspace{20 pt}
\textbf{T}_{i,n}^{3} \leftarrow \sum_{m} \textbf{T}_{i,m}^{2} \times \textbf{W}_{m,n}^{(l)} \nonumber \\
&\textbf{H}_{i,n}^{(l+1)} \leftarrow \sum_{\emptyset} \sigma(\textbf{T}_{i,n}^{3}) \nonumber 
\end{align}
Here, $\hat{\textbf{A}}$ is the adjacency matrix with self-loops added, $\hat{\textbf{D}}$ is the degree matrix, $\textbf{H}^{(l)}$ is the node embedding matrix at layer $l$, and $\textbf{W}^{(l)}$ is the weight matrix at layer $l$. This expression computes the node embeddings for the next layer by: (1) normalizing the adjacency matrix using the degree matrices ($\hat{\textbf{D}}^{-\frac{1}{2}} \hat{\textbf{A}} \hat{\textbf{D}}^{-\frac{1}{2}}$), (2) aggregating neighbor features through this normalized adjacency matrix, and (3) applying a learned transformation through $\textbf{W}^{(l)}$ followed by a non-linear activation function $\sigma$.

We implement a two-layer GCN under transductive setting with the following hyperparameters: Adam optimizer with learning rate $\eta = 0.1$, dropout rate $\gamma = 0.5$, and hidden layer dimension $D = 256$.

All distributed experiments were conducted on servers equipped with Intel\textsuperscript{\textregistered} Xeon\textsuperscript{\textregistered} Silver 4214 CPU, 128GB DDR4 memory, and 1TB SSD storage. The nodes are interconnected via 10 Gbps Ethernet cable. 
Experiments were performed using clusters from 1 to 8 machines. 
Statistics describing the data sets are given in \autoref{table:graph-dataset-statistics}.

\vspace{5 pt}
\noindent
\textit{Results and discussion.}
For large graphs that cannot fit into the memory of a single machine, we use ParMETIS \cite{ParMETIS} to assist with partitioning when running on DGL and AliGraph. The execution times are summarized in \autoref{table:gcn_distributed_running_times}.  To evaluate how accurate the cost model used by the \textsc{SparseEinSum} re-write algorithm is, we execute the best upper-case-lower-case decomposition (\textbf{Opt}), and the second (\textbf{2nd}) and third-best (\textbf{3rd}) as well. 

For the smallest graph (ogbn-arxiv), DGL (running on PyTorch) is  fastest. This is not surprising, as PlinyCompute is built for scalability, not low latency.  For the next larger graph (ogbn-products) the two approaches are about the same.  For the larger two graphs (1B+ edges), the \textsc{SparseEinSum}-generated computation has obvious advantages.  It is the one that can run in all cases---DGL and AliGraph are both plagued by out-of-memory problems, despite manual tuning.  In the cases when DGL can run, it is slower than the \textsc{SparseEinSum}-generated computation by 8.6\% to nearly 40\%.

\begin{figure}[t]
    \centering
    \includegraphics[width=\linewidth]{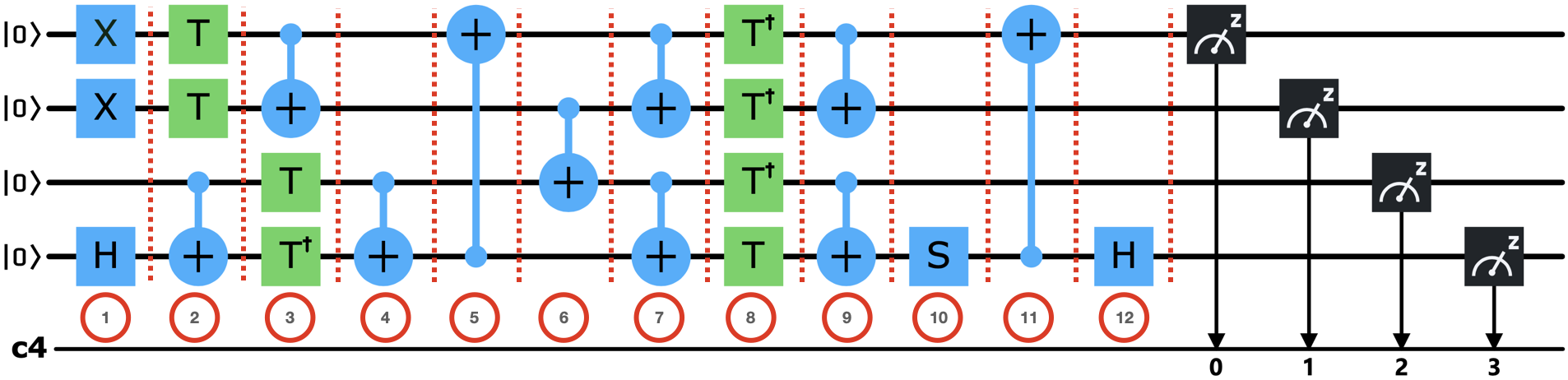}
    \vspace{-15 pt}
    \caption{Example of quantum circuits partitioned into multiple layers from 1 to 12.}
        \vspace{-10 pt}
    \label{fig:quantum_circuit}
\end{figure}

\begin{table}[t]
    \centering
    \begin{tabular}{|l|c|c|c|c|c|}
    \hline
    \textbf{Dataset} & \textbf{Qubits} & \textbf{Depth} & \textbf{Width} & \textbf{Gate Count} \\ \hline
    seca\_n11 & 11 & 37 & 11 & 1275  \\ \hline
    multiplier\_n13 & 13 & 7 & 13 & 496  \\ \hline
    \end{tabular}
    \caption{Dataset statistics for quantum benchmarks.}
    \vspace{-10 pt}
    \label{table:quantum-circuit-dataset}
\end{table}

\begin{table}[t]
\centering
\begin{tabular}{|c|c|c|c|c|}
\hline
\textbf{Circuit} & \textbf{Num} &  \multicolumn{3}{c|}{\textsc{SparseEinSum}}  \\
  \cline{3-5}
 & \textbf{Machines} & \textbf{Opt} & \textbf{2nd Best} & \textbf{3rd Best}  \\
\hline
\multirow{4}{*}{seca\_n11}
      & 1 & 36.124 & \textbf{26.7} & 32.013 \\
      & 2 & \textbf{21.214} & 27.512 & 27.752 \\
      & 4 & \textbf{13.817} & 14.347 & 15.424 \\
      & 8 & \textbf{10.094} & 12.531 & 12.309 \\
\hline
\multirow{4}{*}{multiplier\_n13}
      & 1 & \textbf{102.632} & 105.316 & 116.229 \\
      & 2 & \textbf{55.247} & 64.731 & 71.315 \\
      & 4 & \textbf{36.651} & 42.325 & 47.426 \\
      & 8 & \textbf{21.922} & 25.127 & 30.98 \\
\hline
\end{tabular}
\caption{Time (secs) vs. machines for quantum circuits.}
\label{table:quantum-distributed-runtime-comparison}
\vspace{-20 pt}
\end{table}

\begin{table}[t]
    \centering
    \begin{tabular}{|l|c|c|c|c|c|c|}
    \hline
    \multirow{4}{*}{\textbf{Method}} & \multicolumn{6}{c|}{\textbf{Datasets}} \\
    \cline{2-7}
    & \textbf{Cora} & \textbf{P.} & \textbf{P.Cite} & \textbf{C.Cite} & \textbf{A.} & \textbf{A.}  \vspace{-2 pt} \\
    &  & \textbf{Cora}  & \textbf{Seer} & \textbf{Seer} & \textbf{Photo} & \textbf{Comp} \\
    \hline
    & \multicolumn{6}{c|}{{Sparse Attention Computation}} \\
    \hline
    Hyper & 866 & 901 & 1497 & 40 & 2416 & 7801 \\
    PstGSQL & 8017 & 7990 & 12156 & 619 & 18429 & OOD \\
    SQLite & 22052 & 22084 & TLE & 34875 & TLE & OOD \\
    \textsc{SEinSum} & \textbf{4.5} & \textbf{4.7} & \textbf{8.0} & \textbf{7.7} & \textbf{192} & \textbf{265} \\
    \hline
    & \multicolumn{6}{c|}{{Dense Attention Computation}} \\
    \hline
    
    Hyper & 2194 & 2185 & 3457 & 75 & \textbf{2486} & \textbf{8123} \\
    PstGSQL & 8130 & 8186 & 12432 & 636 & 18464 & OOD \\
    SQLite & 16424 & 15180 & TLE & 39470 & TLE & TLE \\
    \textsc{SEinSum} & \textbf{609} & \textbf{602} & \textbf{950} & \textbf{54} & OOD & OOD \\
    \hline
    \end{tabular}
    \caption{Running time (in seconds) for attention computations. P stands for Planetoid, C: CitationFull, A: Amazon.  OOD is Out of Disk, as the database wrote too much data.}
    \vspace{-10 pt}
    \label{table:gt-runtime-comparison}
\end{table}

\begin{table}[t]
    \centering
    \begin{tabular}{|l|c|c|c|c|c|c|}
    \hline
    \multirow{4}{*}{\textbf{Method}} & \multicolumn{6}{c|}{\textbf{Datasets}} \\
    \cline{2-7}
    & \textbf{Cora} & \textbf{P.} & \textbf{P.Cite} & \textbf{C.Cite} & \textbf{A.} & \textbf{A.}  \vspace{-2 pt} \\
    &  & \textbf{Cora}  & \textbf{Seer} & \textbf{Seer} & \textbf{Photo} & \textbf{Comp} \\
    \hline
    & \multicolumn{6}{c|}{{Graph Conv. Neural Network}} \\
    \hline
    Hyper & 10.7 & 10.8 & \textbf{21.021} & 4.4 & 197.5 & 390.4 \\
    PstGSQL & 16.0 & 16.0 & 30.9 & 7.7 & 436.1 & 830.3 \\
    SQLite & TLE & TLE & TLE & TLE & TLE & TLE  \\
    \textsc{SEinSum} & \textbf{7.3} & \textbf{8.4} & 22.9 & \textbf{3.3} & \textbf{19.3} & \textbf{37.6} \\
    \hline
    \end{tabular}
    \caption{Graph convolutional neural network running times (in seconds). TLE indicates Time Limit Exceeded.}
    \vspace{-18 pt}
    \label{table:gcn-runtime-comparison}
\end{table}

Note that the tensor-relational implementation scales well across different number of machines. \textsc{SparseEinSum}-on-PlinyCompute shows a speed-up of 5.3X, 6.8X, and 5.0X going from one to eight machines on ogbn-products, ogbn-papers100M, and friendster, respectively.  DGL-on-Pytorch fails on the two larger graphs, but has a speedup of only 3.2X on ogbn-products.

\vspace{5 pt}
\noindent
\textbf{Single-machine convolution neural networks.} Single-machine experiments focus on comparing \textsc{SparseEinSum}-generated tensor-relational implementations with equivalent, purely-relational implementations. These are generated using the PortableSQL \EinSum{} compiler \cite{EfficientandPortableSQL}, chosen as it is the standard method in the database literature for converting \EinSum{} to SQL. \textsc{SparseEinSum}-generated tensor-relational computations are run on top of PostgreSQL, again using TACO-generated kernels.   The purely relational implementation is run on SQLite \cite{allen2010introducing}, PostgreSQL \cite{douglas2003postgresql}, and Hyper \cite{kemper2011hyper}. 

\vspace{5 pt}
\noindent
\textit{Results and Discussion.}  Running times are shown in Table \ref{table:gcn-runtime-comparison}. For the four smallest data sets, there is not an appreciable difference between options: the tensor-relational implementation is generally fastest, but it cuts at most 25\% off the running time of ``pure relational + Hyper''. For the larger two data sets, the tensor-relational implementation is ten times faster than ``pure relational + Hyper'', and more than 40 times faster than ``pure relational + Postgres''.  

\begin{table}[t]
    \centering
    \begin{tabular}{|c|c|}
    \hline
    \textbf{UCLC \EinSum{}} & \textbf{Planetoid.CiteSeer}  \\ \hline
    \textbf{$\textbf{T}_{I,k}^{0} \leftarrow \sum_{M} \textbf{X}_{I,M} \times \textbf{W}_{M,k}^{Q}$} & 
    \begin{tabular}{@{}l@{}}$\textbf{X}\texttt{(INT i, INT m, DOUBLE val)}$ \\ $\textbf{W}^{Q}\texttt{(INT m, TENSOR[1024] val)}$ \end{tabular} \\
    \hline
    \textbf{$\textbf{T}_{J,k}^{1} \leftarrow \sum_{N} \textbf{X}_{J,N} \times \textbf{W}_{N,k}^{K}$} & 
    \begin{tabular}{@{}l@{}}$\textbf{X}\texttt{(INT j, INT n, DOUBLE val)}$ \\ $\textbf{W}^{K}\texttt{(INT n, TENSOR[1024] val)}$ \end{tabular} \\
    \hline
    \textbf{$\textbf{T}_{I,J,k}^{2} \leftarrow \sum_{\emptyset} \textbf{T}_{I,k}^{0} \times \textbf{A}_{I,J}$} & 
    \begin{tabular}{@{}l@{}}$\textbf{T}^{0}\texttt{(INT i, TENSOR[1024] val)}$ \\ $\textbf{A}\texttt{(INT i, INT j, DOUBLE val)}$ \end{tabular}  \\
    \hline
    \textbf{$\textbf{T}_{I,J}^{3} \leftarrow \sum_{k} \textbf{T}_{I,J,k}^{2} \times \textbf{T}_{J,k}^{1}$} & 
    \begin{tabular}{@{}l@{}}$\textbf{T}^{2}\texttt{(INT i, INT j,}$ \\ \hspace{40 pt} $\texttt{TENSOR[1024] val)}$ \\ $\textbf{T}^{1}\texttt{(INT j, TENSOR[1024] val)}$ \end{tabular}  \\
    \hline    
    \textbf{$\textbf{Attn}_{I,J} \leftarrow \sum_{\emptyset} \frac{1}{\sqrt{d_k}}(\textbf{T}_{I,J}^{3})$} & 
    $\textbf{T}^{3}\texttt{(INT i, INT j, DOUBLE val)}$  \\ 
    \hline    
    \end{tabular}
    \caption{Upper-case-lower-case Einstein notation produced by the \textsc{SparseEinSum} for Planetoid.CiteSeer.}
    \vspace{-20 pt}
    \label{table:gt-schema}
\end{table}

In Table \ref{table:gcn-schema}, we show the schema generated by the \textsc{SparseEinSum} compiler for two of the input graphs. Interestingly, the computations \emph{are} both implemented quite relationally. However, as one might expect, the exception is that the features and embeddings (which are dense) are implemented as vectors on a per-vertex basis, and the linear transformations are implemented as dense matrices.

\vspace{5 pt}
\noindent
\textbf{Distributed quantum circuit simulation \cite{zhang2021hyquas}.}
We adopt the widely used \texttt{QASMbench} \cite{li2023qasmbench} suite as the quantum circuit simulation benchmark. \texttt{Qiskit} 1.2.4 \cite{wille2019ibm} and \texttt{cuda-quantum} 24.03.0 \cite{kim2023cuda} are used to help with quantum circuit generation and execution.
We select \texttt{seca\_n11} and \texttt{multiplier\_n13} as large scale benchmarks. 
Characteristics for those benchmarks are shown in \autoref{table:quantum-circuit-dataset}.

\begin{table*}[tp]
    \centering
    \begin{tabular}{|c|c|c|c|c|c|}
    \hline
    \small   
    \textbf{Upper-Case-Lower-Case \EinSum{}} & \textbf{CitationFull.CiteSeer} & \textbf{Amazon.Computers} \\ \hline
    \textbf{$\textbf{T}_{I,K}^{0} \leftarrow \sum_{J} \hat{\textbf{D}}_{I,J}\times \hat{\textbf{A}}_{J,K}$} & 
    \begin{tabular}{@{}l@{}}$\hat{\textbf{D}}\texttt{(INT i, INT j, DOUBLE }\texttt{val)}$ \\ $\hat{\textbf{A}}\texttt{(INT j, INT  k,  DOUBLE val)}$ \end{tabular} & 
    \begin{tabular}{@{}l@{}}$\hat{\textbf{D}}\texttt{(INT i, INT j, DOUBLE val)}$ \\ $\hat{\textbf{A}}\texttt{(INT j, INT k,  DOUBLE val)}$ \end{tabular} \\
    \hline
    \textbf{$\textbf{T}_{I,L}^{1} \leftarrow \sum_{K} \textbf{T}_{I,K}^{0}\times \hat{\textbf{D}}_{K,L}$} & 
    \begin{tabular}{@{}l@{}}$\textbf{T}^{0}\texttt{(INT i, INT k, DOUBLE val)}$ \\ $\hat{\textbf{D}}\texttt{(INT k, INT l,  DOUBLE val)}$ \end{tabular} & 
    \begin{tabular}{@{}l@{}}$\textbf{T}^{0}\texttt{(INT i, INT k, DOUBLE val)}$ \\ $\hat{\textbf{D}}\texttt{(INT k, INT l, DOUBLE val)}$ \end{tabular} \\
    \hline
    \textbf{$\textbf{T}_{I,m}^{2} \leftarrow \sum_{L} \textbf{T}_{I,L}^{1} \times \textbf{H}_{L,m}^{(l)}$} & 
    \begin{tabular}{@{}l@{}}$\textbf{T}^{1}\texttt{(INT i, INT l, DOUBLE val)}$ \\ $\textbf{H}^{(l)}\texttt{(INT l,  TENSOR[602] val)}$ \end{tabular} & 
    \begin{tabular}{@{}l@{}}$\textbf{T}^{1}\texttt{(INT i, INT l, DOUBLE val)}$ \\ $\textbf{H}^{(l)}\texttt{(INT l, TENSOR[767] val)}$ \end{tabular} \\
    \hline
    \textbf{$\textbf{T}_{I,n}^{3} \leftarrow \sum_{m} \textbf{T}_{I,m}^{2} \times \textbf{W}_{m,n}^{(l)}$} & 
    \begin{tabular}{@{}l@{}}$\textbf{T}^{2}\texttt{(INT i,  TENSOR[602] val)}$ \\ $\textbf{W}^{(l)}\texttt{ (TENSOR[602][256]  val)}$ \end{tabular} & 
    \begin{tabular}{@{}l@{}}$\textbf{T}^{2}\texttt{(INT i,  TENSOR[767] val)}$ \\ $\textbf{W}^{(l)}\texttt{ (TENSOR[767][256] val)}$ \end{tabular} \\
    \hline 
    \textbf{$\textbf{H}_{I,n}^{(l+1)} \leftarrow \sum_{\emptyset} \sigma(\textbf{T}_{I,n}^{3}) $} & $\textbf{T}^{3}\texttt{(INT i,  TENSOR[256]  val)}$ & $\textbf{T}^{3}\texttt{(INT i,  TENSOR[256] val)}$ \\ \hline
    \end{tabular}
    \caption{Upper-case-lower-case Einstein notation produced by the \textsc{SparseEinSum} compiler for the GCN computation, along with the corresponding relational schemas, for two of the test datasets.}
    \vspace{-10 pt}
    \label{table:gcn-schema}
\end{table*}

\vspace{5 pt}
\noindent
\textit{Computational problem.}
The tensor contraction for quantum gate operations can be written as:
\begin{align}
    \mathbf{\psi}^{\prime}_{i_n} = \sum_{i_{n-1}} \textbf{U}_{i_n,i_{n-1}}^{(n)} \sum_{i_{n-2}} \textbf{U}_{i_{n-1},i_{n-2}}^{(n-1)} \cdots \sum_{i_1} \textbf{U}_{i_2,i_1}^{(2)} \sum_{i_0} \textbf{U}_{i_1,i_0}^{(1)} \mathbf{\psi}_{i_0}
\end{align}
where $\textbf{U}^{(k)}$ represents the unitary gate operation matrix at layer $k$, $\psi$ is the initial quantum state vector, and $\psi^{\prime}$ is the final quantum state vector after applying all gate operations.

Each quantum circuit is partitioned into multiple layers where multiple gates can be executed in parallel within a layer. For each layer, we generate one unitary matrix by tensoring all gates within the layer. The whole simulation process passes qubits vector states into a unitary matrix for each layer one by one as shown in \autoref{fig:quantum_circuit}.

\vspace{5 pt}
\noindent
\textit{Results and Discussion.}
The results for running large-scale benchmarks on a distributed environment are shown in \autoref{table:quantum-distributed-runtime-comparison}.
We evaluate the best, second-best, and third-best decompositions produced, according to the cost model used by the \textsc{SparseEinSum} compiler. In almost all cases, the cost model has ordered the various decompositions correctly. 

Scaling efficiency is reasonable: eight machines is 3.6X faster than one on seca\_n11, and 4.6X faster on multiplier\_n13.
Scaling efficiency is worse than in the case of the GCN, as data movement overhead is high and the computation cost is relatively small.

\vspace{5 pt}
\noindent
\textbf{Single-machine attention computation \cite{ying2021transformers,zhang2024torchgt}.}  Sparse attention resembles the classic dense attention, but it allows an arbitrary graph to indicate which items should be attending to which items.  

\vspace{5 pt}
\noindent
\textit{Computational problem.}  The core computation in a graph transformer is sparse attention computation:
\begin{align}
&\textbf{T}_{i,k}^{0} \leftarrow \sum_{M} \textbf{X}_{i,m} \times \textbf{W}_{m,k}^{Q}; \hspace{20 pt}
\textbf{T}_{j,k}^{1} \leftarrow \sum_{n} \textbf{X}_{j,n} \times \textbf{W}_{n,k}^{K} \nonumber \\
&\textbf{T}_{i,j,k}^{2} \leftarrow \sum_{\emptyset} \textbf{T}_{i,k}^{0} \times \textbf{A}_{i,j}; \hspace{20 pt}
\textbf{T}_{i,j}^{3} \leftarrow \sum_{k} \textbf{T}_{i,j,k}^{2} \times \textbf{T}_{j,k}^{1} \nonumber \\
&\textbf{Attn}_{i,j} \leftarrow \sum_{\emptyset} \frac{1}{\sqrt{d_k}}(\textbf{T}_{i,j}^{3}) \nonumber
\end{align}
where $\textbf{X}$ stores node features, $\textbf{W}_Q$ and $\textbf{W}_K$ are learnable query and key projection matrices respectively, $\textbf{A}$ is the adjacency matrix indicating the graph structure, $d_k$ is the dimension of the key vectors, and $\textbf{Attn}_{i,j}$ gives the attention scores before normalization.
Following prior work \cite{ying2021transformers,zhang2024torchgt}, we evaluate both dense attention (where each node attends to all other nodes) and sparse attention (where nodes only attend to their local neighborhood in the graph).

\vspace{5 pt}
\noindent
\textit{Results and Discussion.} Running time results are given in Table \ref{table:gt-runtime-comparison}. The surprising finding here is that \emph{even though relational computations are inherently sparse}, the pure relational implementations are not much faster when using sparse attention, compared to dense attention. ``Pure relational + Hyper'' best utilizes the sparsity, and is up to 2X as fast in the sparse case compared to the dense case. In comparison, \textsc{SparseEinSum} can be more than 100X as fast when computing sparse attention rather than dense attention, and is 30X to 100X as fast as  ``Pure relational + Hyper'' in the sparse case.

Table \ref{table:gt-schema} shows the generated schema for one of the data sets. The node embeddings are stored as vectors, but the weight matrices are stored into vectors, due to memory limitations. The input data itself is sparse and has been decomposed into purely relational tuples.

\begin{table}[t]
    \centering
    \begin{tabular}{|l|c|c|c|c|c|c|}
    \hline
    \multirow{4}{*}{\textbf{Method}} & \multicolumn{6}{c|}{\textbf{Datasets}} \\
    \cline{2-7}
    & \textbf{Cora} & \textbf{P.} & \textbf{P.Cite} & \textbf{C.Cite} & \textbf{A.} & \textbf{A.}  \vspace{-2 pt} \\
    &  & \textbf{Cora}  & \textbf{Seer} & \textbf{Seer} & \textbf{Photo} & \textbf{Comp} \\
    \hline
    & \multicolumn{6}{c|}{{Sparse Attention Computation}} \\
    \hline
    \textsc{SEinSum} & {4.5} & {4.7} & {8.0} & {7.7} & {192} & {265} \\
    Greedy & 21 & 23 & 52 & 17 & 769 & 884 \\
    \hline
    & \multicolumn{6}{c|}{{Dense Attention Computation}} \\
    \hline
    \textsc{SEinSum} & {609} & {602} & {950} & {54} & OOD & OOD \\
        Greedy & 2198 & 2124 & 4107	& 315 & OOD & OOD \\
    \hline
    \end{tabular}
    \caption{Comparing pure greedy runtime with the full dynamic programming solution for the attention computation.}
    \vspace{-20 pt}
    \label{table:greedy-runtime-comparison}
\end{table}

\subsection{Ablations and Other Experiments}

\textbf{Effectiveness of the dynamic programming.} It is reasonable to ask: Is it necessary to run an expensive and complicated dynamic programming search during the \textsc{SparseEinSum} decomposition?  We evaluate a simpler, greedy search: along a path through the graph of \EinSum{} expressions, greedily pick the decomposition that minimizes the cost; do not remember  decompositions with different output tensor formats that are suboptimal.  We compare greedy to \textsc{SparseEinSum} in Table \ref{table:greedy-runtime-comparison} for attention on one machine.

\begin{table}[t]
    \centering
    \begin{tabular}{|l|c|c|c|c|c|c|}
    \hline
     & \multicolumn{6}{c|}{\textbf{Datasets}} \\
    \cline{2-7}
    Comp- & \textbf{Cora} & \textbf{P.} & \textbf{P.Cite} & \textbf{C.Cite} & \textbf{A.} & \textbf{A.}  \vspace{-2 pt} \\
    onent &  & \textbf{Cora}  & \textbf{Seer} & \textbf{Seer} & \textbf{Photo} & \textbf{Comp} \\
    \hline
    & \multicolumn{6}{c|}{{Sparse Attention Computation}} \\
    \hline
    Exec. & {4.5} & {4.7} & {8.0} & {7.7} & {192} & {265} \\
    Schema & 253 & 96 & 428 & 86 & 344 & 239 \\
    SQL & 46 & 11 & 32 & 2.0 &5.4 & 3.4 \\
    Kernels & 0.04 & 0.04 & 0.04 & 0.08 & 0.04 & 0.04 \\
    \hline
    & \multicolumn{6}{c|}{{Dense Attention Computation}} \\
    \hline
    
   Exec & {609} & {602} & {950} & {54} & OOD & OOD \\
       Schema & 146 & 127 & 533 & 221 & 862 & 1690 \\
    SQL & 51 & 24 & 58 & 44 & 130 & 401 \\
    Kernels & 0.06 & 0.06 &0.06 & 0.06 & 0.06 & 0.06 \\
    \hline
    \end{tabular}
    \caption{Running time for the various system components, attention computation, in seconds.  ``Exec.'' is the actual execution time, ``Schema'' is the time to pre-process the data and generate the schema, ``SQL'' is the time to generate the SQL including the data loading code, and ``Kernels'' is the time to generate the kernels using TACO.}
    \label{table:runtime-coponetns}
    \vspace{-15 pt}
\end{table}

\vspace{5 pt}
\noindent 
\textbf{Robustness to cost model errors.} Is an accurate cost model necessary?  We perturb the cost estimates by multiplying each cost estimate by a Gamma$(\alpha, \theta)$ random variate, where $\alpha \times \theta = 1$ so that on expectation, the cost value is unchanged, but errors are possible.  As we decrease $\alpha$ and increase $\theta$ the variance of the estimate (and hence the potential for error) increases. Results on the single-machine graph neural network computation are given in Table \ref{table:error-tolerance}.  Running time is averaged over ten different perturbed runs. Here we see that with $\alpha = \theta = 1$, there is no difference in the running time---as this is a significant level of noise, it suggests some robustness.  The higher perturbation levels lead to slower times. Interestingly, $\alpha = \frac{1}{3}, \theta = 3$ is worse than $\alpha = \frac{1}{10}, \theta = 10$. 

\begin{table}[t]
    \centering
    \begin{tabular}{|c||c|c|c|c|c|c|}
    \hline
     & \multicolumn{6}{|c|}{\textbf{Perturbation}} \\
\hline
     None & $C_\textrm{xfer}$ & $C_\textrm{xfer}$ & $C_\textrm{krnel}$ & $C_\textrm{krnel}$ & $C_\textrm{fixed}$ & $C_\textrm{fixed}$ \\
     (as-is) & $\times \frac{1}{10}$ & $\times 10$ & $\times \frac{1}{10}$ & $\times 10$ & $\times \frac{1}{10}$ & $\times 10$ \\
    \hline
    \hline
192 & 130 & 253 & 294 & 1027 & 1027 & 305\\
    \hline
    \end{tabular}
    \caption{Effect of perturbing the various cost parameters ($C_\textrm{xfer}, C_\textrm{krnel}, C_\textrm{fixed}$) on the running time for the sparse attention computation on the Amazon.Photo data set.}
    \label{table:error-tolerance}
    \vspace{-20 pt}
\end{table}

\begin{table}[t]
    \centering
    \begin{tabular}{|l|c|c|c|c|c|c|}
    \hline
    \multirow{4}{*}{\textbf{Error}} & \multicolumn{6}{c|}{\textbf{Datasets}} \\
    \cline{2-7}
    & \textbf{Cora} & \textbf{P.} & \textbf{P.Cite} & \textbf{C.Cite} & \textbf{A.} & \textbf{A.}  \vspace{-2 pt} \\
    &  & \textbf{Cora}  & \textbf{Seer} & \textbf{Seer} & \textbf{Photo} & \textbf{Comp} \\
    \hline
    & \multicolumn{6}{c|}{{Graph Conv. Neural Network}} \\
    \hline
    None & {7.3} &{8.4} & 22.9 & {3.3} & {19.3} & {37.6} \\
    $\Gamma(1, 1)$  & {7.2} & {8.3} & {21.5} & {3.4} & {19.5}   & {37.1} \\
    $\Gamma(\frac{1}{3}, 3)$ & {16.1} & {16.0}  & {31.9}  & {8.0} &{435.9} & {839.9} \\
    $\Gamma(\frac{1}{10}, 10)$ & {12.0}  & {11.2}  & {25.1} & {7.4} & {68.0} & {47.4} \\
    \hline
    \end{tabular}
    \caption{Graph neural network running times (in seconds) with erroneous/noisy cardinality estimates. Here,  $\Gamma(\alpha, \theta)$ means each estimate is multiplied by a random number generated using a Gamma$(\alpha, \theta)$ distribution.}
    \vspace{-20 pt}
    \label{table:error-tolerance}
\end{table}

\vspace{5 pt}
\noindent 
\textbf{Robustness to cost parameters.}  We hold two of the three cost parameters ($C_\textrm{xfer}$, $C_\textrm{krnel}$, $C_\textrm{fixed}$) constant, and perturb the third, either cutting it by a factor of ten, or multiplying it by a factor of ten. Single-machine results are given in Table \ref{table:error-tolerance}.  A poor choice in cost parameter can be quite damaging.  Interestingly, cutting $C_\textrm{xfer}$ by a factor of ten actually reduced the running time of the resulting decomposition, suggesting that the $C_\textrm{xfer}$ we chose was not optimal.

\vspace{5 pt}
\noindent 
\textbf{Running time breakdown.} Table \ref{table:runtime-coponetns} compares the execution time for the actual attention computation with the time taken by the decomposition itself.  In addition to the execution time, three other times are given: (1) Schema: time to process the input data, compute the necessary statistics, and run the dynamic programming.  (2) SQL: time to generate the SQL, which is dominated by the time necessary to generate the \texttt{INSERT} \texttt{INTO} statements to load all of the data into PostgreSQL.  (3) Kernels: time to generate the TACO kernels.  We see that these times can be quite significant, though our code is written in Python, and for the single-machine attention experiments, all data are stored in text files.  Further, to ensure portability across engines all data loading is handled using SQL, and is verbose. During deployment or training the attention computation would by run hundreds of times or more, so time  would be amortized across runs.


\section{Related Work}

\texttt{EinSum} was proposed by Einstein \cite{einstein1938gravitational}, and is now supported by machine learning/AI frameworks such as NumPy \cite{numpy}, JAX \cite{jax}, PyTorch \cite{pytorch} and Dask \cite{dask}.  However, upper-case-lower-case \texttt{EinSum}, where \texttt{EinSum} is modified to facilitate sharding specification, is unique.  A general problem in this space is how to specify sharding---see \texttt{https://jax-ml.github.io/scaling-book/sharding/} for an example of such an effort---but  upper-case-lower-case \texttt{EinSum} is unique in its ability to combine a specification of the computation with the way in which the computation is decomposed.

We propose a dynamic programming algorithm for the problem of re-writing a DAG of \EinSum{} expressions (where paths between any connected vertices are unique) to a DAG of upper-case-lower-case \EinSum{}.  Dynamic programming algorithms of this variety have been around for a long time; our algorithm is strongly reminiscent of Felsenstein's algorithm \cite{felsenstein1981evolutionary} from computational biology, and other efforts in sharding of compute graphs for machine learning have used dynamic programming \cite{zheng2022alpa, bourgeois2024eindecomp}. Likewise there exist cost models for sparse tensor computations \cite{ahrens2022autoscheduling, sommer2019mnc}.

There are a number of kernel compilers for tensor computations, such as TVM \cite{chen2018tvm}, Triton \cite{tillet2019triton}, and TACO \cite{taco,10.1145/3276493,kjolstad2019tensor}.  Both TACO and TVM (via SparseTIR \cite{ye2023sparsetir}) provide explicit support for sparsity. Of the various tensor compilers with support for sparsity, Galley \cite{deeds2025galley} is closest to our own proposal, using many ideas from relational system design (such as applying relational methods for sparsity estimation; see Section 7 of this paper). This existing work targets the generation of CPU or GPU kernels or programs, not scalable relational computations that are meant to run on top of a relational system.  The core idea in the \textsc{SparseEinSum} re-write approach is to determine which parts of the computation should be run relationally and which parts should utilize high-performance CPU or GPU kernels generated by these other systems.  In that sense, \textsc{SparseEinSum} sits \emph{on top} of these tools and is a consumer of the kernels that they produce.  In fact, our \textsc{SparseEinSum} prototype uses TACO to generate dense kernels.

There has been other recent work that explores the close relationship between tensors and relations in order to optimize tensor computations. Schleich et al. propose STOREL \cite{schleich2023optimizing}, which accepts a user-supplied storage format for each input tensor, and then optimizes the tensor computation.  In contrast, \textsc{SparseEinSum} accepts a declaratively-specified computation and then optimizes the representation format, choosing how to decompose the tensors into relations so as to optimize performance.  The \textsc{SparseEinSum} approach relies on the underlying relational system to optimize the computation.

In the Duck's Brain \cite{schule2024duck}, Sch{\"u}le et al. are the most recent to argue for the use of relational systems augmented with tensors for modern ML computations---using precisely the sort of tensor-relational representations described in this paper. The Duck's Brain is an example of a \emph{relational platform} on which the output of SparseEinSum-generated tensor-relational computation could be run. The idea of combined tensor and relational processing has been explored in other works \cite{Tensorrelationalalgebra,jankov2021distributed,jankov2019declarative,tang2023auto}, including SystemML \cite{ghoting2011systemml} and SystemDS \cite{boehm2019systemds}, which extended SQL-style operations with matrix primitives.  However, these prior works do not consider how to automatically generate optimized tensor-relational representations. Like in the case of the Duck's Brain, \textsc{SparseEinSum} could be used to generate the decomposition used by these systems.

Another approach explored in the literature is the use of pure relational implementations of ML or AI computations (see, for example Pytond \cite{shahrokhi2024pytond} or Portable SQL \cite {EfficientandPortableSQL} for two recent works). Another recent work proposes an algorithm to switch from dense to sparse dynamically during runtime \cite{staudt2025exploiting}.

\section{Conclusions}
We introduced an algorithm that decomposes tensor computation into equivalent tensor-relational computation, as well as \textsc{SparseEinSum}, a system that takes \EinSum{} computation as input, and compiles computation into upper-case-lower-case Einstein notation. This computation can be implemented on virtually any database system with multidimensional array support. The compilation process incorporates sparsity estimation techniques and automatic cost-based schema optimization, enabling generated tensor-relational computations to achieve significant performance improvements.

\begin{acks}
This research was supported by NSF grants 1918651, 2008240, 2131294, 
and 2212557, NIH CTSA \#UM1TR004906, and US DOT
Tier-1 UTC CYBER-CARE grant \#69A3552348332.
\end{acks}
\bibliographystyle{ACM-Reference-Format}
\bibliography{sample}





\appendix

\end{document}

%% file: ChrisText.tex
\label{sec-EinSum}
The first few paragraphs of this section introduce \EinSum{} and notions such as bounds and label projections, are taken from prior work \cite{bourgeois2024eindecomp}.  

We first define the notion of a \emph{tensor}. We use bold upper-case (for example, $\mathbf{U}$) to denote a tensor. Define the \emph{bound} vector for $\mathbf{U}$, denoted $\mathbf{b}_{\mathbf{U}}$, to be a vector of integers of length $r$. $r$ stands for ``rank''. Matrices are rank-2 tensors. Next, define $\mathcal{I}(\mathbf{b}_{\mathbf{U}})$ to be the set 
$\{0...\mathbf{b}_{\mathbf{U}}[0]  -1\} \times
 \{0...\mathbf{b}_{\mathbf{U}}[1]  -1\} \times ... \times
 \{0...\mathbf{b}_{\mathbf{U}}[r-1]-1\}$.
This is the set of all indices or keys that obey the bound. A tensor $\mathbf{U}$ is then a function from $\mathcal{I}({\mathbf{b}_{\mathbf{U}}})$ to the set of real numbers. 

\vspace{5 pt}
\noindent
\textbf{Simple \EinSum{} examples.}
We start with the classic example: matrix multiplication.  Let $\textbf{U}$ and $\textbf{V}$ be matrices with bounds $[100, 200]$ and $[200, 50]$, respectively.  Then matrix multiplication is written as: $\forall \hspace{2 pt} i,k \in \mathcal{I}\left([100, 50]\right)$:
\begin{align}
\textbf{W}_{i,k} \leftarrow
\sum_{j \in \mathcal{I} \left( [200] \right)}  
  \textbf{U}_{i,j} \times \textbf{V}_{j,k}
\end{align} \label{eqn:matmul}
We often drop the subscript on the aggregation as being un-necessary, as indices from the input tensors that are not used to index into the output tensor are, by definition, aggregated out. The bound vector for the output tensor is likewise implied by the bound vectors of the input, and can also be dropped.

This is an \emph{extended} \EinSum{} notation because we need not necessarily multiply items from the input tensors, and we need not necessarily use summation.  For example, we can compute the maximum squared difference in each row between any two items in that row as:
$\textbf{W}_{i} \leftarrow \textrm{max} (\textbf{U}_{i,j} - \textbf{V}_{i,j})^2.$

\vspace{5 pt}
\noindent
\textbf{General form of \EinSum{}.}
To describe the most general form of \EinSum{}, we define the notion of a \emph{label}, which is a symbol that can be bounded to a value.  We use $\ell_\textbf{U}$ to denote a list (vector) of labels used to index into tensor $\textbf{U}$.

We also need to define the \emph{projection} and \emph{permutation} of a bound vector. Given two lists of labels $\ell_1$ and $\ell_2$, and a bound vector $\textbf{b}$, define $\textbf{b}[\ell_1; \ell_2]$ to be a vector of length $|\ell_1|$, where the $i$th entry is $\textbf{b}[j]$ iff $\ell_1[i]$ = $\ell_2[j]$. As an example, let $\textbf{b} = [ 2, 3, 4 ]$ and let $\ell_1 = [ k, i ]$ and $\ell_2 = [ i, j, k ]$. Then $\textbf{b}[\ell_1; \ell_2] = [ 4, 2 ]$. 

Given this, binary \EinSum{} expressions take the general form:
\begin{align}
\forall \hspace{2 pt} \ell_\textbf{W} \in 
\mathcal{I}\left(\textbf{b}_{\textbf{W}} \right): 
\hspace{2 pt} \textbf{W}_{\ell_\textbf{W}} \leftarrow \hspace{- 7 pt}
\bigoplus_{\ell_\textrm{agg} \in \mathcal{I} \left( \textbf{b}_{\textbf{U} \textbf{V}}[\ell_{\textrm{agg}}; \ell_{\textbf{U} \textbf{V}}] \right)} & \bigotimes \left( \textbf{U}_{\ell_\textbf{U}}, \textbf{V}_{\ell_\textbf{V}} \right)
\end{align} \label{eqn:Einsum}
\noindent Here, $\bigoplus$ is the aggregation operator and $\bigotimes$ is the scalar function applied to joined values (\EinSum{} is  \emph{extended} as it allows for arbitrary $\bigoplus$ and $\bigotimes$ operations). In the above expression, to denote the concatenation of two label lists $\ell_\textbf{U}$ and $\ell_\textbf{V}$, we use $\ell_\textbf{UV}$. $\textbf{b}_{\textbf{U} \textbf{V}}$ similarly denotes the concatenation of two bound vectors.

Consider a more complicated \EinSum{} expression. Assume that we have two tensors $\textbf{U}$ and $\textbf{V}$ with bound vectors $\textbf{b}_\textbf{U} = [ 10, 100, 20 ]$ and  $\textbf{b}_\textbf{V} = [ 100, 20, 2000 ]$.  We wish to transpose $\textbf{U}$ to obtain a tensor with bound $[ 20, 10, 100 ]$, then transpose $\textbf{V}$ to obtain a new tensor with bound $[ 20, 100, 2000 ]$, and do a batch matrix multiply for the two resulting tensors, and then sum out the batch dimension.

In \EinSum{}, this is expressed in the single expression:
\begin{align*}
\forall i,k \in \mathcal{I}([ 10, 2000 ]), \textbf{W}_{i,k} \leftarrow
\sum_{b,j \in \mathcal{I} \left( [ 20, 100 ] \right)}  
  \textbf{U}_{i,j,b} \times \textbf{V}_{j,b,k}
\end{align*}
\noindent Considering the general form, we have $\ell_\textbf{U} = [ i,j,b ]$, $\ell_\textbf{V} = [ j,b,k ]$, $\ell_\textrm{agg} = [ b,j ]$ and $\textbf{b}_\textbf{UV} = [ 10, 100, 20, 100, 20, 2000 ]$.  The bound vector for the aggregation is computed as $\textbf{b}_{\textbf{U} \textbf{V}}[\ell_{\textrm{agg}}; \ell_{\textbf{U} \textbf{V}}]$. How? $\ell_\textrm{agg}$ has two labels: $b$ and $j$. As $b$ occupies the third (and fifth) position in $\ell_{\textbf{U} \textbf{V}}$, and $j$ occupies the second (and fourth) position in $\ell_{\textbf{U} \textbf{V}}$, we select the third (or fifth) item in $\textbf{b}_\textbf{UV}$, and the second (or fourth) item results in $\textbf{b}_\textbf{UV}$. This results in the bound vector $[ 20, 100 ]$.  

\vspace{5 pt}
\noindent
\textbf{Multi-Layer Perceptron in \EinSum{}.}
Using  \EinSum{}, we are able to specify very complex computations, including modern transformers, by composing individual \EinSum{} operators together into a directed, acyclic graph of operators. For a simple example of such a DAG, the following specifies a fully-connected neural network with two weight matrices, $\textbf{W}_{i,j}^{(1)}$ and $\textbf{W}_{i,j}^{(2)}$ used to connect the input to a hidden layer, and then a softmax:
\begin{align}
    &\textbf{A}_{i} \leftarrow \sum_j  \textbf{W}_{i,j}^{(1)} \times \textbf{X}_{j}; \hspace{.3 cm}
    \textbf{B}_{i} \leftarrow \sum_{\emptyset} \textrm{ReLU} ( \textbf{A}_{i}) \hspace{.3 cm}
    \textbf{C}_{i} \leftarrow \sum_j  \textbf{W}_{i,j}^{(2)} \times \textbf{B}_{j};  \nonumber \\
    &\textbf{D}_{\emptyset} \leftarrow \sum_i \exp (\textbf{C}_{i}); \hspace{.3 cm}
    \textbf{E}_{i} \leftarrow \sum_{\emptyset} \frac{\exp(\textbf{C}_{i})} {\textbf{D}_{\emptyset}} \nonumber 
\end{align}

\section{Sparse \EinSum{}}
\label{sec-sparse}

\subsection{Why Worry About Sparsity?}

\EinSum{} is agnostic as to the implementation of the underlying tensors, but this implementation can have a significant practical effect on the efficiency of real-life calculations, especially if the underlying tensors are sparse.  Consider the following two matrices:
$$ \textbf{U} \equiv \begin{bmatrix}
1.4 & 2.2 & 0 & 2.1 \\
0 & 0 & 0 & 0\\
1.4 & 0 & 1.1 & 0\\
0 & 0 & 0 & 0 
\end{bmatrix} \textbf{V} \equiv \begin{bmatrix}
3.2 & 0 & 1.3 & 0 \\
0 & 0 & 0.6 & 0\\
0 & 0 & 1.2 & 0\\
1.2 & 0 & 2.1 & 0 
\end{bmatrix}$$ 
If we store these matrices in a classic, dense format such as row-major or column-major, most classical implementations of the matrix multiplication specified by the \EinSum{} $\textbf{W}_{i,k} \leftarrow
\sum \textbf{U}_{i,j} \times \textbf{V}_{j,k}$, would perform 64 pairwise scalar multiplications.  However, we could store these two matrices \emph{relationally}, and have a much more efficient implementation of the matrix multiplication, at least in terms of the number of multiplications required.  Consider representing \textbf{U} as a set of vectors, where any vectors containing only zero values are not represented explicitly:
$$ \bar{\textbf{U}} \equiv  \left\{ \left( 0, \begin{bmatrix}  1.4 & 2.2 & 0 & 2.1 \end{bmatrix} \right),  \left(  2, \begin{bmatrix}  1.4 & 0 & 1.1 & 0 \end{bmatrix} \right) \right\}$$
$\bar{\textbf{U}}$ is implemented as a relation with schema \texttt{(i, valU)} where \texttt{valU} is of type \texttt{vector[4]}. \textbf{V} is implemented as a set of vectors:
\begin{equation}
    \bar{\textbf{V}} \equiv  \left\{ \left( 0, \begin{bmatrix}  3.2 \\ 0 \\ 0 \\ 1.2 \end{bmatrix} \right),  \left( 2, \begin{bmatrix}  1.3 \\ 0.6 \\ 1.2 \\ 2.1 \end{bmatrix} \right) \right\} \label{output-W} 
\end{equation} with schema \texttt{(k, valV)}.  The same matrix multiplication can now be implemented as a simple SQL query that performs a join with a dot product over the cross product of the two relations:

\begin{SQL}
SELECT U.i, V.k, 
    dot_product (U.valU, V.valV) AS valW
FROM $\bar{\textbf{U}}$ AS U, $\bar{\textbf{V}}$ AS V
\end{SQL}

Note that the result is a fully relational (fully decomposed) implementation of the resulting matrix, with schema  \texttt{(i, k, valW)}:  $$ \textbf{W} \equiv  \left\{ \left( 0, 0, 7 \right),  \left( 0, 2, 7.55 \right), \left( 2, 0, 4.48 \right), \left( 2, 2, 3.14 \right)\right\}$$
Crucially, there are only 16 pairwise scalar multiplications required to perform this implementation, four of which will be invoked by the call to the \texttt{dot\_product()} operation. Thus, only 25\% as many scalar multiplications are required as the traditional way.  

\subsection{Promoting and Demoting Labels}

For any relational implementation, there will be an overhead associated with pushing each tuple through the system, and this overhead may render the original matrix multiplication the best option. Still, there is clearly \emph{some} level of sparsity for which this decomposed representation is preferred due to the lower computational load.

For any tensor, many decompositions into \emph{tensor relations} of this form are possible.
Consider an \EinSum{} expression over tensor $\textbf{U}$ with bound $\textbf{b}_\textbf{U}$ and label list $\ell_\textbf{U}$. We describe a \emph{decomposition} of $\textbf{U}$ by a partitioning of $\ell_\textbf{U}$ into two lists, $\up{U}$ and $\down{U}$:

\begin{enumerate}
\item The labels in $\up{U}$ are \emph{promoted} so that they index into the resulting relation.

\item The labels in $\down{U}$ are \emph{demoted} so that they index into the tensors within the relation. 
\end{enumerate}

\noindent
The decomposition of $\textbf{U}$ induced by $\up{U}$ and $\down{U}$ is a database relation with schema 
$$\bar{\textbf{U}}\texttt{(INT }\up{U}[1]\texttt{, INT } \up{U}[2]\texttt{, ...,} \texttt{ TENSOR }\texttt{valU)}$$
In each tuple, $\texttt{valU}$ is a tensor with bound  $\textbf{b}_\textbf{U}[\down{U}; \ell_\textbf{U}]$, and the $i$th attribute $\up{U}[i]$ in the relation takes the integer values from $0$ through $\textbf{b}_\textbf{U}[\up{U}; \ell_\textbf{U}][i] - 1$.  Crucially, assuming that $\oplus$ and $\otimes$ form a semiring, if $\texttt{valU}$ is filled only with the special zero value of the semiring, the corresponding tuple can be eliminated from the resulting tensor relation, facilitating sparsity.

The set of all decompositions of a tensor $\textbf{U}$ in an \EinSum{} expression defines an equivalence class of all \emph{tensor-relational} implementations for $\textbf{U}$ in the expression, in that all of the tensor-relational implementations describe same mapping from $\mathcal{I}(\textbf{b}_\textbf{U})$ to the real numbers.  Let $\bar{\textbf{U}}$ be a decomposition of $\textbf{U}$, defined by $\up{U}$ and $\down{U}$.  Then, for any index 
$\textbf{i} \in \mathcal{I}(\textbf{b}_\textbf{U})$, the following relational query returns a single tuple containing a scalar with the value at $\textbf{U}_\textbf{i}$:

\begin{SQL}
SELECT (ISNULL (valU$_{\textbf{i}[\downn{U}; \ell_\textbf{U}]}$, zero))
FROM (SELECT valU FROM $\bar{\textbf{U}}$ WHERE $\up{U} = \textbf{i}[\up{U}; \ell_\textbf{U}]$) 
     LEFT OUTER JOIN ZERO
\end{SQL}

Here, we assume a relation \texttt{ZERO(zero)} that contains a single tuple with the zero value for the semiring. The \texttt{WHERE} clause of the inner query uses the set of labels promoted to the relational part of the representation to select the tuple corresponding to \textbf{i}, and then the outer \texttt{SELECT} clause uses the remaining labels to index into the tensor \texttt{valU} to obtain the desired scalar.  If the inner query returns no results, then \texttt{valU} will be \texttt{NULL} and the query result will be the zero value for the associated semiring. 

For a concrete example of this, re-consider \textbf{V}, its (possible) tensor-relational representation $\bar{\textbf{V}}$ from this section, and the  \EinSum{} $\textbf{W}_{i,k} \leftarrow
\sum \textbf{U}_{i,j} \times \textbf{V}_{j,k}$.  In this case, $\bar{\textbf{V}}$ corresponds to the decomposition $\up{V} = k$,  $\down{V} = j$.  Consider the indexing vector $\textbf{i} = [3, 2]$.  In this case, $\textbf{i}[\down{V}; \ell_\textbf{V}] = \textbf{i}[j; jk] = 3$ and  $\textbf{i}[\up{V}; \ell_\textbf{V}] = \textbf{i}[k; jk] = 2$.  Thus, to obtain $\textbf{V}[3, 2]$ in  $\bar{\textbf{V}}$, we would use the query:

\begin{SQL}
SELECT (ISNULL (valV$_3$, zero))
FROM (SELECT valV FROM $\bar{\textbf{V}}$ WHERE k = 2) 
     LEFT OUTER JOIN ZERO
\end{SQL}

\subsection{``Upper-Case-Lower-Case'' Notation}

Rather than explicitly listing $\up{U}$ and $\down{U}$, we suggest the convention that in an \EinSum{} expression over tensor relations, any labels in \emph{upper-case} are in $\up{U}$, and any labels in \emph{lower-case} are in $\down{U}$. This defines the \emph{upper-case-lower-case} notation. So, consider the following \EinSum{} expression for a tensor-relational matrix multiplication:
$$\textbf{W}_{i,K} \leftarrow
\sum \textbf{U}_{i,J} \times \textbf{V}_{J,K}$$
In this case, labels $J, K$ are used to index into relations, and label $i$ into tensors.  So this corresponds to the implementation where \textbf{U} is represented as the set:
$$ \left\{ \left(0, \begin{bmatrix}
1.4 \\
0 \\
1.4 \\
0 
\end{bmatrix} \right),
\left(1, \begin{bmatrix}
 2.2 \\
 0 \\
 0 \\
 0 
\end{bmatrix} \right),
\left(2, \begin{bmatrix}
 0 \\
0 \\
 1.1 \\
0 
\end{bmatrix}  \right),
\left(3, \begin{bmatrix}
2.1 \\
 0\\
 0\\
 0 
\end{bmatrix} \right) \right\} $$
and \textbf{V} is represented as the set:
$$\{
(0, 0, 3.2), (0, 2, 1.3), (1, 2, 0.6), (2, 2, 1.2),
(3, 0, 1.2), (3, 2, 2.1)
\}$$ 
while the output \textbf{W} is again represented as a set of column vectors, as in Equation \ref{output-W}.

For another example, consider the decomposition from the Introduction (the chain of two multiplications used to implement message passing in a graph neural network). Following the convention that those indexes promoted to a relational representation are given in upper case, this can be specified via the \EinSum{}:
\begin{align}
\textbf{T}_{K,i} \leftarrow
\sum \textbf{X}_{i,J} \times \textbf{Y}_{J,K}; \textrm{   }
\textbf{U}_{I,k} \leftarrow
\sum \textbf{T}_{I,j} \times \textbf{Z}_{j,k} \nonumber
\end{align}

\section{Decomposing \EinSum{}}

We can decompose any binary \EinSum{} expression into an equivalent computation over tensor relations.  This equivalence gives a roadmap to implement tensor-relational \EinSum{} over sparse inputs relationally.
Using the convention of the last section of the paper, $\bar{\textbf{U}}$ and  $\bar{\textbf{V}}$ are used to denote the tensor-relational decompositions of $\textbf{U}$ and  $\textbf{V}$, respectively. Now, assume that we have an associated \emph{kernel function} $\mathcal{K}$.   This is a function that implements the \EinSum{}:
\begin{align}
\textbf{W}_{\downn{W}} \leftarrow 
\bigoplus \bigotimes \left( \textbf{U}_{\downn{U}}, \textbf{V}_{\downn{V}} \right) \nonumber
\end{align}
Re-write equation \ref{eqn:Einsum} as:
\begin{align}
\textbf{W}_{\ell_\textbf{W}} \leftarrow 
\bigoplus_{\downagg{}, \upagg{}} \bigotimes \left( \left( \sigma_\upp{U} \left(\bar{\textbf{U}} \right) \right)_\downn{U},  \left( \sigma_\upp{V} \left(\bar{\textbf{V}} \right) \right)_\downn{V} \right) \label{eqn_tr}
\end{align}
A bit of additional notation is introduced here. Given a binding of the labels in $\up{U}$ to values in the vector $\textbf{i}$ during the execution of the \EinSum{} expression, $\sigma_\upp{U} \left(\bar{\textbf{U}} \right)$ represents the application of relational selection equivalent to running the query:

\begin{SQL}
SELECT ISNULL (valU, zero)
FROM (SELECT valU FROM $\bar{\textbf{U}}$ WHERE $\upp{U} = \textbf{i}$) 
     LEFT OUTER JOIN ZERO
\end{SQL}

Note that \texttt{zero} is a tensor, filled with zeros, with bound $\textbf{b}[\down{U}; \ell_{\textbf{U}}]$. Further, $\upagg$ and $\downagg$ represent the aggregation labels in the original expression partitioned into two lists, depending upon whether they appear in  $\up{U}, \up{V}$ or in $\down{U}, \down{V}$.

Given this, Equation \ref{eqn_tr} can be re-written as:
\begin{align}
\textbf{W}_{\ell_\textbf{W}} \leftarrow 
\bigoplus_{\upagg{}} \left(  \left( \sigma_{\upp{W}, \upagg{}} \left( \Join \left(\bar{\textbf{U}}, \bar{\textbf{V}},
\texttt{valW} \leftarrow \mathcal{K} \left(\texttt{valU}, \texttt{valV} \right)
\right) \right) \right)_{\downn{W}} \right) \nonumber
\end{align}
Here, $\Join \left(\bar{\textbf{U}}, \bar{\textbf{V}}, \texttt{valW} \leftarrow \mathcal{K}(.) \right)$ performs a natural join of the relations $\bar{\textbf{U}}$ and  $\bar{\textbf{V}}$ and computes additional attribute \texttt{valW} using the kernel function $\mathcal{K}$.
Define $\bigoplus_{\upagg{}}$ relationally so that it aggregates the tensors stored in \texttt{valW} by applying the $\oplus$ operator pairwise to each item in a pair of tensors, then this last expression can be re-written as $\textbf{W}_{\ell_\textbf{W}} \leftarrow$
\begin{align}
\left(  \sigma_{\upp{W}} \left( \bigoplus_{\upagg{}} \left( \Join \left(\bar{\textbf{U}}, \bar{\textbf{V}},
\texttt{valW} \leftarrow \mathcal{K} \left(\texttt{valU}, \texttt{valV} \right) 
\right) \right) \right) \right)_{\downn{W}}   \label{eqn-TRA}
\end{align}
This suggests a tensor-relational implementation for any binary \EinSum{} expression over tensor relations:

\begin{SQL}
SELECT $\up{W}$, SUM ($\mathcal{K} \left(\texttt{valU}, \texttt{valV}  \right)$) AS valW 
FROM $\bar{\textbf{U}}$ NATURAL JOIN $\bar{\textbf{V}}$
GROUP BY $\up{W}$
\end{SQL}

\section{Compiling \EinSum{} into Tensor-Relational SQL}
\label{sec-compiling}

This suggests a framework for compiling a DAG of \EinSum{} expressions into tensor-relational algebra or tensor-relational SQL.

\begin{enumerate}
\item For each vertex (\EinSum{}) expression in the DAG, re-write the \EinSum{} into upper-case-lower-case \EinSum{}.
\item For each resulting expression, generate SQL according to Equation \ref{eqn-TRA}. That is, generate a \texttt{SELECT}-\texttt{FROM}-\texttt{WHERE}-\texttt{GROUP BY} query where (a) the attributes chosen in the \texttt{SELECT} are those labels in $\up{W}$ as well as ``\texttt{SUM}$(\mathcal{K} \left(\texttt{valU}, \texttt{valV} \right))$ as \texttt{valW}''; (b) the \texttt{WHERE} clause performs a natural join of the two input relations; (c) we group by those attributes corresponding to labels in $\up{W}$.
\item If there are mismatches in tensor-relational decomposition across decomposed \EinSum{} expressions, generate tensor-relational algebra/tensor-relational SQL to change representation.
\end{enumerate}

The last point bears some discussion.  Imagine, for example, that we generated the following decomposition:
\begin{align}
\textbf{T}_{K,i} \leftarrow
\sum \textbf{X}_{i,J} \times \textbf{Y}_{J,K} \hspace{15 pt} 
\textbf{U}_{i,k} \leftarrow
\sum \textbf{T}_{i,J} \times \textbf{Z}_{J,k} \nonumber
\end{align}
SQL for the first expression would be:

\begin{SQL}
SELECT K, SUM ($\mathcal{K}_1 \left(\texttt{valX}, \texttt{valY} \right)$) as valT
FROM $\bar{\textbf{X}}$ NATURAL JOIN $\bar{\textbf{Y}}$
GROUP BY K
\end{SQL}

SQL for the second expression would be:

\begin{SQL}
SELECT SUM($\mathcal{K}_2 \left(\texttt{valT}, \texttt{valZ} \right)$) as valU
FROM $\bar{\textbf{T}}$ NATURAL JOIN $\bar{\textbf{Z}}$
\end{SQL}

The output decomposition of \textbf{T} from the first \EinSum{} (a set of row vectors, as the first [row] index is upper-case) does not match the input decomposition of \textbf{T} to the second \EinSum{} (a set of column vectors, as the second [column] index is upper case). Thus, code must be generated to repartition \textbf{T} between the operations.
Assume the output of the first \EinSum{} expression---partitioned according to $\textbf{T}_{K,i}$, named as $\bar{\textbf{T}}_\textrm{in}$, and we want to repartition according to $\textbf{T}_{i,J}$ to form $\bar{\textbf{T}}$.
We would first decompose in the first dimension:

\begin{SQL}
CREATE VIEW $\bar{\textbf{T}}_\textrm{int}$(i, j, valT) AS
SELECT $\bar{\textbf{T}}_\textrm{in}$.k, A.index, $\bar{\textbf{T}}_\textrm{in}$.valT[A.index]  
FROM ALLINTS (0, $\textbf{b}_\textbf{T}[1]$) AS A, $\bar{\textbf{T}}_\textrm{in}$
\end{SQL}

In this code, $\bar{\textbf{T}}_\textrm{in}$\texttt{.valT[A.index]} extracts the sub-tensor (here a scalar) at position \texttt{A.index} from $\bar{\textbf{T}}_\textrm{in}$\texttt{.valT}.  \texttt{ALLINTS} is a table function returning all values in the specified range.  And then we re-compose on the second dimension:

\begin{SQL}
CREATE VIEW  $\bar{\textbf{T}}$(j, valT) AS
SELECT $\bar{\textbf{T}}_\textrm{int}$.j, STACK ($\bar{\textbf{T}}_\textrm{int}$.valT, $\bar{\textbf{T}}_\textrm{int}$.i, 0, $\textbf{b}_\textbf{T}$[0])
FROM $\bar{\textbf{T}}_\textrm{int}$
GROUP BY $\bar{\textbf{T}}_\textrm{int}$.j
\end{SQL}

Here, \texttt{STACK} is an aggregation function that can be used to stack scalars (or tensors) to create a higher-dimensional tensor. It accepts: (1) the value to stack, (2) the position of this value in the higher-dimensional tensor, (3) the new dimension to create, and (4) the maximum value of this dimension. 

In the general case, the repartition from tensor relation $\bar{\textbf{T}}_\textrm{in}$ to $\bar{\textbf{T}}$ happens in two steps. First, there is a decomposition of the result of the first \EinSum{} expression so that it is partitioned on \emph{all} of the labels in either $\up{T}_\textrm{in}$ or  $\up{T}$.  Then there is an ``un-partitioning'' or aggregation to obtain $\bar{\textbf{T}}$.  Note that if $\up{T}_\textrm{in}$ is a subset of $\up{T}$, the aggregation can be skipped, and if $\up{T}$ is a subset of $\up{T}_\textrm{in}$ the decomposition can be skipped.  If $\up{T}_\textrm{in} = \up{T}$, then the entire repartition can be skipped, as the partitionings are the same.

\section{The Sparse \EinSum{} Cost Model}
\label{sec-costmodel}
This leaves two final questions, as our goal is to find the best decomposition.  (1) How do we cost a given decomposition in sparsity-aware fashion?
(2) How to efficiently search the space of all decompositions?  In this section, we consider the first question.

\subsection{Estimating Sizes Under Sparsity}

When developing a cost model, the first question we must address is:
given a (potentially sparse) tensor $\textbf{U}$ with label set $\ell_\textbf{U}$ and a tensor-relational decomposition $\bar{\textbf{U}}$ defined by $\up{U}$ and $\down{U}$, how many tuples are in $\bar{\textbf{U}}$?  That is, if we take a tensor and decompose it into a tensor relation, can we estimate how many tuples will be contained in the relation? We define this quantity using the notation $T(\bar{\textbf{U}})$. 

We begin by defining two statistics that are sufficient to accurately estimate $T(\bar{\textbf{U}})$. First, for a label $l$ in $\ell_\textbf{U}$, let $V(l, \textbf{U})$ denote the number of sub-tensors induced by $l$ that have at least one non-zero entry, if we were to use $\up{U} = \{l\}$. In other words, $V(l, \textbf{U})$ is the answer to the question: If we were to decompose $\textbf{U}$ by representing label $l$ relationally, how many tuples would we have?  

To be precise, let $\ell'_\textbf{U}$ denote $\ell_\textbf{U}$ with $l$ removed.  Then $V(l, \textbf{U})$ can be computed using the following \EinSum{} expressions:
\begin{align} \textbf{V}_l \leftarrow \sum_{\ell'_\textbf{U}} \texttt{if} (\textbf{U}_{\ell_\textbf{U}} == 0, 0, 1) \nonumber \\
 V(l, \textbf{U}) \leftarrow \sum_{l} \texttt{if} (\textbf{V}_{l} == 0, 0, 1) \nonumber
\end{align}
\noindent Here, the function ``\texttt{if (bool, val1, val2})'' returns \texttt{val1} if \texttt{bool} is true, and \texttt{val2} otherwise.

Note that $V(l, \textbf{U})$ is in some sense equivalent to the notion of the number of distinct values for an attribute that is commonly used in relational query optimization. Why? Imagine that $\textbf{U}$ were represented purely relationally---that is, we perform a tensor-relational decomposition of $\textbf{U}$ into $\bar{\textbf{U}}$ so that $\up{U} = \ell_{\textbf{U}}$ and $\down{U} = \{\}$, and remove any tuples storing the scalar value zero. Now consider attribute $l$. If $l$ has $d$ distinct values in $\bar{\textbf{U}}$, it means that were we to group all of the tuples having the same $l$ value into a sub-tensor of $\textbf{U}$, there would be exactly $d$ sub-tensors, and none of those would be composed entirely of zeros (any sub-tensors composed purely of zeros would not contribute any tuples to $\bar{\textbf{U}}$, and hence would not contribute to the count $d$). Thus, $V(\textbf{U}, l)$ would be $d.$

Further, assume we have $T({\textbf{U}})$, which is the number of non-zero entries in tensor $\textbf{U}$.  Again, this is analogous to the number of tuples in a relation. If we perform a pure relational decomposition of $\textbf{U}$ and remove tuples storing zero values, then the number of non-zero entries in the tensor $\textbf{U}$ would be the same as the number of tuples in such a pure relational decomposition. 

Let $n(\up{U})$ denote the number of possible tuples in some tensor-relational decomposition $\bar{\textbf{U}}$ for $\textbf{U}$, induced by the promoted label set $\up{U}$  This can be computed as: 
$$n(\up{U}) = \prod_{l \in \upp{U}} V(l, \textbf{U})$$
Then the number of tuples in the tensor-relational decomposition $\bar{\textbf{U}}$ induced by the set of promoted labels $\up{U}$ can be estimated as:
\begin{align}T(\up{U}) = n(\up{U}) \left( 1 - \textrm{exp} \left( {\frac{-T({\textbf{U}})}{n(\up{U})}}\right)\right)\label{eqn-count} \end{align}
\noindent The intuition is that there are $n(\up{U})$ possible tuples in the tensor relation $\bar{\textbf{U}}$. $T(\up{U})$ estimates the number of tuples that have tensors with at least one non-zero value, if the $T({\textbf{U}})$ non-zero values are spread uniformly-at-random throughout the tensor relation.

\subsection{The Cost Model}


When implementing a generic, binary \EinSum{} expression of the form of Equation \ref{eqn:Einsum} tensor-relationally, there are two operations, a join and an aggregation.  There is a third operation---repartition---that may be necessary to change tensor-relational decompositions between the output of one operation, and the input of the next.  We consider how to cost each of them.

\vspace{5 pt}
\noindent
\textbf{Costing the join.} The set of labels participating in the relational join $\ell_{\textrm{join}}$ is the list of labels in both $\up{U}$ and $\up{V}$.  The number of tuples resulting from the tensor-relational join can be estimated as:
\begin{align}
T_{\textrm{join}}(\up{U}, \up{V}) = \frac{T(\up{U}) T(\up{V})}{\prod_{l \in \ell_{\textrm{join}}} \textrm{max} (V(\textbf{U}, l), V(\textbf{V}, l))} \nonumber
\end{align}
This is an application of the standard, textbook estimator for the number of tuples resulting from a join. Note that if there are no labels common to $\up{U}$ and $\up{V}$ so that $\ell_{\textrm{join}}$ is empty, it is assumed that the denominator takes the value $1$. 
Then the cost of the tensor-relational join is estimated as:
\begin{align}
C_{\textrm{join}}(\up{U}, \up{V}) &= \nonumber \\T_{\textrm{join}}(\up{U}, \up{V}) &\times \big[ \big(\texttt{sizeof} (\bar{\textbf{U}}) + \texttt{sizeof} (\bar{\textbf{V}})) C_{\textrm{xfer}} + C_{\textrm{krnel}} + C_\textrm{fixed} \big] \nonumber
\end{align}
\noindent Here, \texttt{sizeof} computes the size of a tuple from the tensor relation, in bytes, $C_{\textrm{xfer}}$ is the per-byte cost to move a tuple from machine to machine (assuming a distributed environment), $C_{\textrm{krnel}}$ is the cost to invoke the kernel $\mathcal{K}$ (typically this is the number of floating-point operations required by $\mathcal{K}$, times a constant), and  $C_{\textrm{fixed}}$ is a fixed, per-tuple processing cost.  Overall, the cost is then: (a) the cost to move tuples from both inputs into the join; plus (b) the cost to run the kernel, plus (b) a fixed cost per tuple.  

A rule-of-thumb is to choose the ratio $C_{\textrm{xfer}}$ by checking how long it takes to transfer a byte from machine to machine (or from disk to machine in a single-machine environment), to choose $C_{\textrm{krnel}}$ by checking how long it takes to run a floating point computation on the execution platform, and to choose $C_{\textrm{fixed}}$ by running a single-machine computation that produces and aggregates a large number of tuples (say, by running a cross product of two relations and aggregating the result) and dividing the runtime by the number of tuples produced to obtain the per-tuple overhead.

\vspace{5 pt}
\noindent
\textbf{Costing the aggregation.} 
The number of tuples resulting from the subsequent aggregation can be estimated as 
\begin{align}
T_\textrm{agg} (\up{U}, \up{V}) &= \textrm{min} \Bigg( \frac{1}{2} T_{\textrm{join}}(\up{U}, \up{V}), \nonumber \\
&\prod_{l \in \upagg{}} 
\begin{cases}
\textrm{if } l \in \up{U} \wedge  l \in \up{V}: \textrm{min} (V(\textbf{U}, l), V(\textbf{V}, l))  \\
\textrm{else if } l \in \up{U} : V(\textbf{U}, l) \\
\textrm{else} : V(\textbf{V}, l)
\end{cases} \vspace{-25 pt}
\Bigg) \nonumber
\end{align}
\noindent 
This is a variation on the standard, textbook estimator for the number of tuples resulting from an aggregation.  Note the three cases in the second option of the ``min''. The number of distinct values for each promoted grouping label is either (a) the minimum of the number of distinct values from both $\textbf{U}$ and $\textbf{V}$ if the label is a join attribute; (b) the number of distinct values from $\textbf{U}$ if the label comes only from $\textbf{U}$; (c) the  number of distinct values from $\textbf{V}$, otherwise.
Then the cost of the aggregation is estimated as:
\begin{align}
C_{\textrm{agg}}(\up{U}, \up{V}) &= \nonumber \\  (T_{\textrm{join}}(\up{U}, \up{V}) &- T_{\textrm{agg}}(\up{U}, \up{V})) \times \big[ \texttt{sizeof} (\bar{\textbf{W}})  C_{\textrm{xfer}} + C_{\textrm{add}} + C_\textrm{fixed} \big] \nonumber
\end{align}
This formula mirrors the formula for $C_{\textrm{join}}$, the one difference being that the output of the aggregation is $T_{\textrm{agg}}(\up{U}, \up{V})$ ``aggregation groups''; reducing the $\frac{T_{\textrm{join}}(\upp{U}, \upp{V})}{T_{\textrm{agg}}(\upp{U}, \upp{V})}$ tuples in a aggregation group down to a single tuple requires moving and adding $\frac{T_{\textrm{join}}(\upp{U}, \upp{V})}{T_{\textrm{agg}}(\upp{U}, \upp{V})} - 1$ tuples, or $T_{\textrm{join}}(\up{U}, \up{V}) - T_{\textrm{agg}}(\up{U}, \up{V})$ moves-and-adds overall.

\vspace{5 pt}
\noindent
\textbf{Costing the repartition.} 
Finally, there is the cost to repartition tensor relation $\bar{\textbf{W}}$ when the output partition of the expression producing $\bar{\textbf{W}}$ does not match the input to the consumer of $\bar{\textbf{W}}$.   

Let $\up{W}_\textrm{in}$ be the set of promoted labels in the input to the repartition, and let $\up{W}$ be the set of promoted labels in the target, consuming \EinSum{} expression.

As illustrated in the SQL given in the prior section, the typical implementation of a repartition first decomposes the result of the first \EinSum{} expression so that it is partitioned on \emph{all} of the labels in either  $\up{W}_\textrm{in}$ or  $\up{W}$ (while both $\up{W}_\textrm{in}$ and  $\up{W}$ are lists and not sets, we abuse notation and denote the list containing all labels in both as $\up{W}_\textrm{in} \cup \up{W}$). The number of tuples in the resulting tensor relation can be estimated using $T(\up{W}_\textrm{in} \cup \up{W})$.  Note that this quantity cannot be less than either $T(\up{W})$ or $T(\up{W}_\textrm{in})$, as the intermediate decomposition uses a set of promoted labels that is a superset of both the input and output set of promoted labels.

Then it is possible to compute an expression for the cost $C_\textrm{repart}$:
\begin{align}C_\textrm{repart}&(\up{W}_\textrm{in}, \up{W}) = \nonumber \\ &\big(T(\up{W}_\textrm{in} \cup \up{W}) -  T(\up{W}_\textrm{in})  \big) \big( \texttt{sizeof} (\bar{\textbf{W}}_\textrm{int}) C_\textrm{xfer} + C_\textrm{fixed} \big) + \nonumber \\  &\big(T(\up{W}_\textrm{in} \cup \up{W}) - T(\up{W}) \big) \big( \texttt{sizeof} (\bar{\textbf{W}}_\textrm{int}) C_\textrm{xfer} + C_\textrm{fixed} \big)  \nonumber \\
&= \big(2T(\up{W}_\textrm{in} \cup \up{W}) - T(\up{W}_\textrm{in}) - T(\up{W}) \big) \times \nonumber \\ &\big( \texttt{sizeof} (\bar{\textbf{W}}_\textrm{int}) C_\textrm{xfer} + C_\textrm{fixed} \big) \nonumber
\end{align}
\noindent $C_\textrm{repart}$ has two parts.  First is the cost of creating and processing the $T(\up{W}_\textrm{in} \cup \up{W}) -  T(\up{W}_\textrm{in})$ new tuples during the first, decomposition step.  And second is aggregating those tuples into $T(\up{W})$ inputs to the next operation; this requires moving around $T(\up{W}_\textrm{in} \cup \up{W}) - T(\up{W})$ intermediate tuples.

\begin{algorithm}[tb]
   \caption{\textsc{ComputeMinCostForTree}}
   \label{algorithm:opt}
\begin{algorithmic}[1]
\STATE \textbf{topologically} sort the vertices of the input \EinSum{} graph $G$
\STATE \textbf{initialize} lookup cost table $C$ so that for any vertex $v$ with no inputs, $C[v, \up{W}] = 0$ forall $ \up{W} \in \ell_\textbf{W}$, where $\ell_\textbf{W}$ is the list of labels for the tensor $\textbf{W}$ produced by $v$
\FOR{each vertex $v \in V$, in sorted order}
    \STATE \% input tensors always have cost 0
    \IF{$v$ has no inputs}
        \STATE \textbf{continue}
    \ENDIF
    \STATE find $v_\textbf{U}$ s.t. $(v_\textbf{U}, v, \texttt{left}) \in E$
    \STATE find $v_\textbf{V}$ s.t. $(v_\textbf{V}, v, \texttt{right}) \in E$
    \STATE $cost_\textrm{min} \leftarrow \infty$
    \STATE \textbf{extract} $\ell_\textbf{U}$, $\ell_\textbf{V}$, $\ell_\textbf{W}$ from vertex $v$ 
    \STATE \% Here it will consider every output decomp for $\textbf{W}$
    \FOR{each subset $\up{W} \in \ell_{\textbf{W}}$}
       \STATE \% $\up{U}$ is \emph{consistent} with $\up{W}$ if, for each $l$ in $(\ell_{\textbf{U}} \cap \ell_{\textbf{W}})$, 
       \STATE \% $l$ is in $\up{U}$ iff it is also in $\up{W}$

       \FOR{each subset $\up{U} \in \ell_{\textbf{U}}$ that is consistent with $\up{W}$} 
          \STATE $cost_\textbf{U} \leftarrow \min\limits_{\upp{U}_\textrm{in}} C_\textrm{repart} (\up{U}_\textrm{in}, \up{U}) + C[v_\textbf{U}, \up{U}_\textrm{in}]$
          \FOR{each subset $\up{V} \in \ell_{\textbf{V}}$ that is consistent with $\up{W}$}
              \STATE $cost_\textbf{V} \leftarrow \min\limits_{\upp{V}_\textrm{in}} C_\textrm{repart} 
              (\up{V}_\textrm{in}, \up{V}) + C[v_\textbf{V}, \up{V}_\textrm{in}]$
              \STATE $cost_\textbf{W} \leftarrow cost_\textbf{U} + cost_\textbf{V} +$ \\ \hspace{20 pt} $C_\textrm{join} (\up{U}, \up{V}) + C_\textrm{agg} (\up{U}, \up{V})$
              \IF{$cost_\textbf{W} < cost_\textrm{min}$}
                 \STATE $cost_\textrm{min} \leftarrow cost_\textbf{W}$         
              \ENDIF
          \ENDFOR
       \ENDFOR
       \STATE \% remember the low cost to compute $\textbf{W}$ via decomp $\up{W}$
       \STATE $C[v, \up{W}] \leftarrow cost_\textrm{min}$
    \ENDFOR
\ENDFOR
\STATE \textbf{let} $v$ be the last vertex in $V$
\STATE \textbf{return} $\min\limits_{\upp{W}} C[v, \up{W}]$
\end{algorithmic}
\end{algorithm}

\section{Optimizing the Decomposition}
\label{sec-optimizing}
We now define a dynamic programming algorithm that computes the lowest cost, sparsity-aware, decomposition of a DAG of \EinSum{} expressions. This algorithm re-writes each expression in the DAG into upper-case-lower-case notation.  Combined with the results from Section 5, this allows such an \EinSum{} program to be compiled into a high-performance, tensor-relational computation.

\subsection{Dynamic Programming}


Consider a directed graph $G = \langle V, E \rangle$, where each vertex in the vertex set $V$ represents an \EinSum{} operation. $V$ has two types of vertices: (1) those with no inputs (these correspond to input tensors) and (2) those with two inputs (these correspond to binary \EinSum{} expressions).  For our optimization algorithm, the important features of each \EinSum{} operation are the sets of labels $\ell_\textbf{U}$, $\ell_\textbf{V}$, $\ell_\textbf{W}$ present in the expression, as well as the bound vectors of the input and output tensors.  These will allow us to make use of the cost models from the previous section.  

Each directed edge in $E$ is a triple consisting of (1) the vertex/\EinSum{} expression producing a tensor, (2) the vertex/\EinSum{} expression consuming that same tensor, and (3) a constant that is either \texttt{left} or \texttt{right} indicating if this tensor is the left or right input into the consuming vertex.

The dynamic programming algorithm works its way through the graph, considering vertices in topologically-sorted order. As it does so, it populates a lookup table $C$. For a vertex $v$ producing a tensor with label list $\ell_\textbf{W}$, $C[v, \up{W}]$ will be computed for every possible set of promoted labels $ \up{W}$---this is all possible subsets of $\ell_\textbf{W}$. $C[v, \up{W}]$ stores the minimum possible cost to produce tensor $\textbf{W}$---considering all of the \EinSum{} expressions necessary to compute  $\textbf{W}$---subject to the constraint that the tensor-relational decomposition of the output tensor is described by $\up{W}$. 

It is possible to populate $C$ in a single pass through the graph via a dynamic programming algorithm because as long as each tensor is consumed by exactly one \EinSum{} expression, we do not care about the details of how the output associated with a vertex $v$ is computed; we only care about (a) the output decomposition resulting from $v$, and (b) the cost to compute $v$.  As long as $C$ stores the lowest cost to compute all decompositions of an \EinSum{} expression corresponding to vertex $v$, to optimize the decomposition of a consumer of $v$, we need only look at all entries of the form $C[v, .]$. 

The dynamic programming algorithm is given as Algorithm \ref{algorithm:opt}. For each vertex $v$ there is a triply-nested loop. At the outer level, we are computing the optimal cost for every possible tensor-relational decomposition of the output of $v$, described by the set of promoted labels $\up{W}$.  The goal is to compute the lowest cost to produce $\up{W}$ for $v$. The two inner loops consider all possible input decompositions that are \emph{consistent} with $\up{W}$.  If $\up{W}$ promotes a label $l$ so that it is handled relationally, then both inputs must promote that label. By considering all of these consistent input combinations, the optimal value for $C[v, \up{W}]$ is computed.

\begin{algorithm}[tb]
   \caption{\textsc{GetStats}}
   \label{alg-stats}
\begin{algorithmic}[1]
\STATE \textbf{compile} input \EinSum{} DAG into SQL over relations storing only scalars---that is, let $\up{U} = \ell_{\textbf{U}}$ and  $\down{U} = \{\}$ for each tensor $\textbf{U}$, and compute statistics assuming that all tuples storing zero-valued scalars are removed from the resulting relations
\STATE \textbf{compile} purely-relational SQL into relational algebra using classical methods
\STATE \textbf{cost} relational algebra using classical methods, to compute/estimate the number of tuples for each input and intermediate relation; for relation $\bar{\textbf{U}}$ corresponding to tensor $\textbf{U}$, let $T(\textbf{U})$ denote this tuple count
\STATE \textbf{cost} relational algebra using classical methods, to compute/estimate the number of distinct values for each attribute; for attribute $l$ and relation $\bar{\textbf{U}}$, let $V(\textbf{U}, l)$ denote this value count

\end{algorithmic}
\end{algorithm}

\subsection{Additional Considerations}

\noindent
\textbf{Pre-computing the required statistics.}
In practice, running Algorithm \autoref{algorithm:opt} requires that for each tensor $\textbf{U}$, we have $T(\textbf{U})$ (the number of non-zero entries in $\textbf{U}$), and that we have $V(\textbf{U}, l)$ for each label $l$ (the number of non-zero tensors induced by label $l$).
Obtaining these statistics can be achieved by running the \textsc{GetStats} algorithm: Algorithm \autoref{alg-stats}. At the highest level, this algorithm relies on generating SQL corresponding to a fully-relational version of the input \EinSum{} program, and then using classical relational costing methods to compute the required statistics. 

\vspace{5 pt}
\noindent
\textbf{When results are consumed more than once.}
The prior dynamic programming algorithm cannot work in this case, as the algorithm relies on the set of entries $C[v, .]$ fully describing all information necessary to optimally use the \EinSum{} expression corresponding to vertex $v$. However, data in machine learning computations (both input and intermediate) are often used multiple times.  For example, during backpropagation, intermediate data are used both during the forward and backward passes. In both the graph convolutional neural network and the attention computation in our experiments, input data are used in two \EinSum{} expressions. If $v$ is used more than once, an optimal algorithm would need to worry about both uses of $v$, and co-optimize those. 

The approach we implement is to decompose an \EinSum{} program into a set of ``trees''. That is, we first choose a subgraph of program $G$ where no \EinSum{} result is used more than once (in general, this subgraph should be ``maximal'' in the sense that if we can add another vertex and it is still the case that no vertex is used more than once, we should do so). Then, we optimize this subgraph, using Algorithm \ref{algorithm:opt}. At this point, the tensor-relational decomposition of each tensor that has been optimized is treated as being fixed.  This process is then repeated, and another subgraph of program $G$ where no \EinSum{} result is used more than once is chosen and optimized, until all decompositions have been chosen.

This will not arrive at a globally optimal solution, and it could be the case that some tensor-relational decompositions that are used twice are not chosen so as to be globally optimal.  However, as the cost of this is only a sub-optimal, additional repartition, the expected cost of this non-optimality will hopefully be inconsequential.

\vspace{5 pt}
\noindent
\textbf{Memory constraints.}
Often, in terms of performance, the optimal schema design is to use only massive tensors, and not to decompose the tensors at all. In practice, this will lead to memory problems. In our implementation, we address this by constraining the search to reflect memory limitations---maximum sizes for the total bytes required for the inputs and output of a kernel call, for example.

\vspace{5 pt}
\noindent
\textbf{Non-Binary \EinSum{} expressions.}
All of the algorithms and cost models described thus far in the paper can easily be extended to unary \EinSum{} expressions, and we do not consider this issue further.

Higher-degree \EinSum{} expressions are more challenging since the contraction path is critical for performance.  While rare in ML computations, they do occur, particularly in other application domains.  Our suggested tactic is to use optimization libraries like \texttt{opt\_einsum} \cite{opteinsum} to compute the contraction path and then break such expressions as a series of binary \EinSum{} expressions.  More sophisticated approaches are a topic for future work.

%% file: sample.bib
@String{Computing = "Computing" }

@String{Springer = "Springer-Verlag" }

@article{EfficientandPortableSQL,
author = {Blacher, Mark and Klaus, Julien and Staudt, Christoph and Laue, S\"{o}ren and Leis, Viktor and Giesen, Joachim},
title = {Efficient and Portable Einstein Summation in SQL},
year = {2023},
issue_date = {June 2023},
publisher = {Association for Computing Machinery},
address = {New York, NY, USA},
volume = {1},
number = {2},
url = {https://doi.org/10.1145/3589266},
doi = {10.1145/3589266},
journal = {Proc. ACM Manag. Data},
month = {jun},
articleno = {121},
numpages = {19}
}

@article{kipf2016semi,
  title={Semi-supervised classification with graph convolutional networks},
  author={Kipf, Thomas N and Welling, Max},
  journal={arXiv preprint arXiv:1609.02907},
  year={2016}
}

@misc{QASMBench,
  title = {QASMBench},
  howpublished = {\url{https://github.com/pnnl/QASMBench}}
}

@misc{ParMETIS,
  title = {ParMETIS},
  howpublished = {\url{https://github.com/KarypisLab/ParMETIS}}
}

@inproceedings{zou2018plinycompute,
  title={PlinyCompute: A platform for high-performance, distributed, data-intensive tool development},
  author={Zou, Jia and Barnett, R Matthew and Lorido-Botran, Tania and Luo, Shangyu and Monroy, Carlos and Sikdar, Sourav and Teymourian, Kia and Yuan, Binhang and Jermaine, Chris},
  booktitle={Proceedings of the 2018 International Conference on Management of Data},
  pages={1189--1204},
  year={2018}
}

@book{douglas2003postgresql,
  title={PostgreSQL: a comprehensive guide to building, programming, and administering PostgresSQL databases},
  author={Douglas, Korry and Douglas, Susan},
  year={2003},
  publisher={SAMS publishing}
}

@article{li2023qasmbench,
  title={Qasmbench: A low-level quantum benchmark suite for nisq evaluation and simulation},
  author={Li, Ang and Stein, Samuel and Krishnamoorthy, Sriram and Ang, James},
  journal={ACM Transactions on Quantum Computing},
  volume={4},
  number={2},
  pages={1--26},
  year={2023},
  publisher={ACM New York, NY}
}

@inproceedings{kim2023cuda,
  title={Cuda quantum: The platform for integrated quantum-classical computing},
  author={Kim, Jin-Sung and McCaskey, Alex and Heim, Bettina and Modani, Manish and Stanwyck, Sam and Costa, Timothy},
  booktitle={2023 60th ACM/IEEE Design Automation Conference (DAC)},
  pages={1--4},
  year={2023},
  organization={IEEE}
}

@inproceedings{wille2019ibm,
  title={IBM’s Qiskit tool chain: Working with and developing for real quantum computers},
  author={Wille, Robert and Van Meter, Rod and Naveh, Yehuda},
  booktitle={2019 Design, Automation \& Test in Europe Conference \& Exhibition (DATE)},
  pages={1234--1240},
  year={2019},
  organization={IEEE}
}

@inproceedings{chen2018tvm,
  title={$\{$TVM$\}$: An automated $\{$End-to-End$\}$ optimizing compiler for deep learning},
  author={Chen, Tianqi and Moreau, Thierry and Jiang, Ziheng and Zheng, Lianmin and Yan, Eddie and Shen, Haichen and Cowan, Meghan and Wang, Leyuan and Hu, Yuwei and Ceze, Luis and others},
  booktitle={13th USENIX Symposium on Operating Systems Design and Implementation (OSDI 18)},
  pages={578--594},
  year={2018}
}

@inproceedings{shahrokhi2024pytond,
  title={Pytond: Efficient python data science on the shoulders of databases},
  author={Shahrokhi, Hesam and Kaboli, Amirali and Ghorbani, Mahdi and Shaikhha, Amir},
  booktitle={2024 IEEE 40th International Conference on Data Engineering (ICDE)},
  pages={423--435},
  year={2024},
  organization={IEEE}
}

@article{Tensorrelationalalgebra,
author = {Yuan, Binhang and Jankov, Dimitrije and Zou, Jia and Tang, Yuxin and Bourgeois, Daniel and Jermaine, Chris},
title = {Tensor relational algebra for distributed machine learning system design},
year = {2021},
issue_date = {April 2021},
publisher = {VLDB Endowment},
volume = {14},
number = {8},
issn = {2150-8097},
url = {https://doi.org/10.14778/3457390.3457399},
doi = {10.14778/3457390.3457399},
journal = {Proc. VLDB Endow.},
month = apr,
pages = {1338–1350},
numpages = {13}
}

@article{xiang2025cutespmm,
  title={cuTeSpMM: Accelerating Sparse-Dense Matrix Multiplication using GPU Tensor Cores},
  author={Xiang, Lizhi and Asudeh, Omid and Sabin, Gerald and Sukumaran-Rajam, Aravind and Sadayappan, P},
  journal={arXiv preprint arXiv:2504.06443},
  year={2025}
}

@article{schule2024duck,
  title={The Duck’s Brain: Training and Inference of Neural Networks within Database Engines},
  author={Sch{\"u}le, Maximilian and Neumann, Thomas and Kemper, Alfons},
  journal={Datenbank-Spektrum},
  volume={24},
  number={3},
  pages={209--221},
  year={2024},
  publisher={Springer}
}

@article{deeds2025galley,
  title={Galley: Modern Query Optimization for Sparse Tensor Programs},
  author={Deeds, Kyle and Ahrens, Willow and Balazinska, Magdalena and Suciu, Dan},
  journal={Proceedings of the ACM on Management of Data},
  volume={3},
  number={3},
  pages={1--24},
  year={2025},
  publisher={ACM New York, NY, USA}
}

@article{schleich2023optimizing,
  title={Optimizing tensor programs on flexible storage},
  author={Schleich, Maximilian and Shaikhha, Amir and Suciu, Dan},
  journal={Proceedings of the ACM on Management of Data},
  volume={1},
  number={1},
  pages={1--27},
  year={2023},
  publisher={ACM New York, NY, USA}
}

@inproceedings{kemper2011hyper,
  title={HyPer: A hybrid OLTP\&OLAP main memory database system based on virtual memory snapshots},
  author={Kemper, Alfons and Neumann, Thomas},
  booktitle={2011 IEEE 27th International Conference on Data Engineering},
  pages={195--206},
  year={2011},
  organization={IEEE}
}

@article{allen2010introducing,
  title={Introducing SQLite},
  author={Allen, Grant and Owens, Mike and Allen, Grant and Owens, Mike},
  journal={The Definitive Guide to SQLite},
  pages={1--16},
  year={2010},
  publisher={Springer}
}

@inproceedings{yang2016revisiting,
  title={Revisiting semi-supervised learning with graph embeddings},
  author={Yang, Zhilin and Cohen, William and Salakhudinov, Ruslan},
  booktitle={International conference on machine learning},
  pages={40--48},
  year={2016},
  organization={PMLR}
}

@article{bojchevski2017deep,
  title={Deep gaussian embedding of graphs: Unsupervised inductive learning via ranking},
  author={Bojchevski, Aleksandar and G{\"u}nnemann, Stephan},
  journal={arXiv preprint arXiv:1707.03815},
  year={2017}
}

@article{shchur2018pitfalls,
  title={Pitfalls of graph neural network evaluation},
  author={Shchur, Oleksandr and Mumme, Maximilian and Bojchevski, Aleksandar and G{\"u}nnemann, Stephan},
  journal={arXiv preprint arXiv:1811.05868},
  year={2018}
}

@article{hu2020open,
  title={Open graph benchmark: Datasets for machine learning on graphs},
  author={Hu, Weihua and Fey, Matthias and Zitnik, Marinka and Dong, Yuxiao and Ren, Hongyu and Liu, Bowen and Catasta, Michele and Leskovec, Jure},
  journal={Advances in neural information processing systems},
  volume={33},
  pages={22118--22133},
  year={2020}
}

@article{ying2021transformers,
  title={Do transformers really perform badly for graph representation?},
  author={Ying, Chengxuan and Cai, Tianle and Luo, Shengjie and Zheng, Shuxin and Ke, Guolin and He, Di and Shen, Yanming and Liu, Tie-Yan},
  journal={Advances in neural information processing systems},
  volume={34},
  pages={28877--28888},
  year={2021}
}

@inproceedings{zhang2024torchgt,
  title={TorchGT: A Holistic System for Large-Scale Graph Transformer Training},
  author={Zhang, Meng and Sun, Jie and Hu, Qinghao and Sun, Peng and Wang, Zeke and Wen, Yonggang and Zhang, Tianwei},
  booktitle={SC24: International Conference for High Performance Computing, Networking, Storage and Analysis},
  pages={1--17},
  year={2024},
  organization={IEEE}
}

@inproceedings{tillet2019triton,
  title={Triton: an intermediate language and compiler for tiled neural network computations},
  author={Tillet, Philippe and Kung, Hsiang-Tsung and Cox, David},
  booktitle={Proceedings of the 3rd ACM SIGPLAN International Workshop on Machine Learning and Programming Languages},
  pages={10--19},
  year={2019}
}

@article{taco,
  author = {Kjolstad, Fredrik and Kamil, Shoaib and Chou, Stephen and Lugato, David and Amarasinghe, Saman},
  title = {The tensor algebra compiler},
  year = {2017},
  issue_date = {October 2017},
  publisher = {Association for Computing Machinery},
  address = {New York, NY, USA},
  volume = {1},
  number = {OOPSLA},
  url = {https://doi.org/10.1145/3133901},
  doi = {10.1145/3133901},
  journal = {Proc. ACM Program. Lang.},
  month = oct,
  articleno = {77},
  numpages = {29}
}

@article{numpy,
  title={Array programming with NumPy},
  author={Harris, Charles R and Millman, K Jarrod and Van Der Walt, St{\'e}fan J and Gommers, Ralf and Virtanen, Pauli and Cournapeau, David and Wieser, Eric and Taylor, Julian and Berg, Sebastian and Smith, Nathaniel J and others},
  journal={Nature},
  volume={585},
  number={7825},
  pages={357--362},
  year={2020},
  publisher={Nature Publishing Group UK London}
}

@article{pytorch,
  title={Pytorch: An imperative style, high-performance deep learning library},
  author={Paszke, A},
  journal={arXiv preprint arXiv:1912.01703},
  year={2019}
}

@article{jax,
  title={Compiling machine learning programs via high-level tracing},
  author={Frostig, Roy and Johnson, Matthew James and Leary, Chris},
  journal={Systems for Machine Learning},
  volume={4},
  number={9},
  year={2018}
}

@article{opteinsum,
  title={Opt$\backslash$\_einsum-a python package for optimizing contraction order for einsum-like expressions},
  author={Daniel, G and Gray, Johnnie and others},
  journal={Journal of Open Source Software},
  volume={3},
  number={26},
  pages={753},
  year={2018}
}

@book{dask,
  title={Data science with Python and Dask},
  author={Daniel, Jesse},
  year={2019},
  publisher={Simon and Schuster}
}

@inproceedings{ghoting2011systemml,
  title={SystemML: Declarative machine learning on MapReduce},
  author={Ghoting, Amol and Krishnamurthy, Rajasekar and Pednault, Edwin and Reinwald, Berthold and Sindhwani, Vikas and Tatikonda, Shirish and Tian, Yuanyuan and Vaithyanathan, Shivakumar},
  booktitle={2011 IEEE 27th International conference on data engineering},
  pages={231--242},
  year={2011},
  organization={IEEE}
}

@article{boehm2019systemds,
  title={SystemDS: A declarative machine learning system for the end-to-end data science lifecycle},
  author={Boehm, Matthias and Antonov, Iulian and Baunsgaard, Sebastian and Dokter, Mark and Ginth{\"o}r, Robert and Innerebner, Kevin and Klezin, Florijan and Lindstaedt, Stefanie and Phani, Arnab and Rath, Benjamin and others},
  journal={arXiv preprint arXiv:1909.02976},
  year={2019}
}

@article{jankov2021distributed,
  title={Distributed numerical and machine learning computations via two-phase execution of aggregated join trees},
  author={Jankov, Dimitrije and Yuan, Binhang and Luo, Shangyu and Jermaine, Chris},
  journal={Proceedings of the VLDB Endowment},
  volume={14},
  number={7},
  year={2021}
}

@article{jankov2019declarative,
  title={Declarative recursive computation on an rdbms, or, why you should use a database for distributed machine learning},
  author={Jankov, Dimitrije and Luo, Shangyu and Yuan, Binhang and Cai, Zhuhua and Zou, Jia and Jermaine, Chris and Gao, Zekai J},
  journal={arXiv preprint arXiv:1904.11121},
  year={2019}
}

@inproceedings{tang2023auto,
  title={Auto-differentiation of relational computations for very large scale machine learning},
  author={Tang, Yuxin and Ding, Zhimin and Jankov, Dimitrije and Yuan, Binhang and Bourgeois, Daniel and Jermaine, Chris},
  booktitle={International Conference on Machine Learning},
  pages={33581--33598},
  year={2023},
  organization={PMLR}
}

@inproceedings{staudt2025exploiting,
  title={Exploiting Dynamic Sparsity in Einsum},
  author={Staudt, Christoph and Blacher, Mark and Hoffmann, Tim and Kasche, Kaspar and Beyersdorff, Olaf and Giesen, Joachim},
  booktitle={The Thirty-ninth Annual Conference on Neural Information Processing Systems},
  year={2025}
}

@article{10.1145/3276493,
author = {Chou, Stephen and Kjolstad, Fredrik and Amarasinghe, Saman},
title = {Format abstraction for sparse tensor algebra compilers},
year = {2018},
issue_date = {November 2018},
publisher = {Association for Computing Machinery},
address = {New York, NY, USA},
volume = {2},
number = {OOPSLA},
url = {https://doi.org/10.1145/3276493},
doi = {10.1145/3276493},
journal = {Proc. ACM Program. Lang.},
month = oct,
articleno = {123},
numpages = {30}
}

@inproceedings{kjolstad2019tensor,
  title={Tensor algebra compilation with workspaces},
  author={Kjolstad, Fredrik and Ahrens, Willow and Kamil, Shoaib and Amarasinghe, Saman},
  booktitle={2019 IEEE/ACM International Symposium on Code Generation and Optimization (CGO)},
  pages={180--192},
  year={2019},
  organization={IEEE}
}

@inproceedings{ye2023sparsetir,
  title={Sparsetir: Composable abstractions for sparse compilation in deep learning},
  author={Ye, Zihao and Lai, Ruihang and Shao, Junru and Chen, Tianqi and Ceze, Luis},
  booktitle={Proceedings of the 28th ACM International Conference on Architectural Support for Programming Languages and Operating Systems, Volume 3},
  pages={660--678},
  year={2023}
}

@inproceedings{zhang2021hyquas,
  title={HyQuas: hybrid partitioner based quantum circuit simulation system on GPU},
  author={Zhang, Chen and Song, Zeyu and Wang, Haojie and Rong, Kaiyuan and Zhai, Jidong},
  booktitle={Proceedings of the 35th ACM International Conference on Supercomputing},
  pages={443--454},
  year={2021}
}

@inproceedings{boyd2004friendster,
  title={Friendster and publicly articulated social networking},
  author={Boyd, Danah Michele},
  booktitle={CHI'04 extended abstracts on Human factors in computing systems},
  pages={1279--1282},
  year={2004}
}

@inproceedings{sommer2019mnc,
  title={Mnc: Structure-exploiting sparsity estimation for matrix expressions},
  author={Sommer, Johanna and Boehm, Matthias and Evfimievski, Alexandre V and Reinwald, Berthold and Haas, Peter J},
  booktitle={Proceedings of the 2019 International Conference on Management of Data},
  pages={1607--1623},
  year={2019}
}

@inproceedings{ahrens2022autoscheduling,
  title={Autoscheduling for sparse tensor algebra with an asymptotic cost model},
  author={Ahrens, Willow and Kjolstad, Fredrik and Amarasinghe, Saman},
  booktitle={Proceedings of the 43rd ACM SIGPLAN International Conference on Programming Language Design and Implementation},
  pages={269--285},
  year={2022}
}

@article{bourgeois2024eindecomp,
  title={EinDecomp: Decomposition of Declaratively-Specified Machine Learning and Numerical Computations for Parallel Execution},
  author={Bourgeois, Daniel and Ding, Zhimin and Jankov, Dimitrije and Li, Jiehui and Sleem, Mahmoud and Tang, Yuxin and Yao, Jiawen and Yao, Xinyu and Jermaine, Chris},
  journal={arXiv preprint arXiv:2410.02682},
  year={2024}
}

@inproceedings{zheng2022alpa,
  title={Alpa: Automating inter-and $\{$Intra-Operator$\}$ parallelism for distributed deep learning},
  author={Zheng, Lianmin and Li, Zhuohan and Zhang, Hao and Zhuang, Yonghao and Chen, Zhifeng and Huang, Yanping and Wang, Yida and Xu, Yuanzhong and Zhuo, Danyang and Xing, Eric P and others},
  booktitle={16th USENIX Symposium on Operating Systems Design and Implementation (OSDI 22)},
  pages={559--578},
  year={2022}
}

@article{felsenstein1981evolutionary,
  title={Evolutionary trees from DNA sequences: a maximum likelihood approach},
  author={Felsenstein, Joseph},
  journal={Journal of molecular evolution},
  volume={17},
  number={6},
  pages={368--376},
  year={1981},
  publisher={Springer}
}

@article{einstein1938gravitational,
  title={The gravitational equations and the problem of motion},
  author={Einstein, Albert and Infeld, Leopold and Hoffmann, Banesh},
  journal={Annals of mathematics},
  pages={65--100},
  year={1938},
  publisher={JSTOR}
}

@article{zhou2020graph,
  title={Graph neural networks: A review of methods and applications},
  author={Zhou, Jie and Cui, Ganqu and Hu, Shengding and Zhang, Zhengyan and Yang, Cheng and Liu, Zhiyuan and Wang, Lifeng and Li, Changcheng and Sun, Maosong},
  journal={AI open},
  volume={1},
  pages={57--81},
  year={2020},
  publisher={Elsevier}
}

@article{wu2020comprehensive,
  title={A comprehensive survey on graph neural networks},
  author={Wu, Zonghan and Pan, Shirui and Chen, Fengwen and Long, Guodong and Zhang, Chengqi and Yu, Philip S},
  journal={IEEE transactions on neural networks and learning systems},
  volume={32},
  number={1},
  pages={4--24},
  year={2020},
  publisher={IEEE}
}

@misc{wang2020deepgraphlibrarygraphcentric,
      title={Deep Graph Library: A Graph-Centric, Highly-Performant Package for Graph Neural Networks}, 
      author={Minjie Wang and Da Zheng and Zihao Ye and Quan Gan and Mufei Li and Xiang Song and Jinjing Zhou and Chao Ma and Lingfan Yu and Yu Gai and Tianjun Xiao and Tong He and George Karypis and Jinyang Li and Zheng Zhang},
      year={2020},
      eprint={1909.01315},
      archivePrefix={arXiv},
      primaryClass={cs.LG},
      url={https://arxiv.org/abs/1909.01315}, 
}

@misc{zhu2019aligraphcomprehensivegraphneural,
      title={AliGraph: A Comprehensive Graph Neural Network Platform}, 
      author={Rong Zhu and Kun Zhao and Hongxia Yang and Wei Lin and Chang Zhou and Baole Ai and Yong Li and Jingren Zhou},
      year={2019},
      eprint={1902.08730},
      archivePrefix={arXiv},
      primaryClass={cs.DC},
      url={https://arxiv.org/abs/1902.08730}, 
}

@article{barr1991einstein,
  title={The Einstein summation notation},
  author={Barr, Alan H},
  journal={An Introduction to Physically Based Modeling (Course Notes 19), pages E},
  volume={1},
  pages={57},
  year={1991}
}
